\documentclass[twocolumn,twocolappendix]{aastex63}

\received{}
\revised{}
\accepted{}
\shorttitle{DESI $z\gtrsim 5$ Quasar Survey}
\shortauthors{Yang et al.}

\usepackage{threeparttable}

\begin{document}

\title{DESI $z \gtrsim 5$ Quasar Survey. I. A First Sample of 400 New Quasars at $z \sim 4.7 - 6.6$}

\correspondingauthor{Jinyi Yang}
\email{jinyiyang@email.arizona.edu}

\author[0000-0001-5287-4242]{Jinyi Yang}
\altaffiliation{Strittmatter Fellow}
\affiliation{Steward Observatory, University of Arizona, 933 N Cherry Avenue, Tucson, AZ 85721, USA}

\author[0000-0003-3310-0131]{Xiaohui Fan}
\affiliation{Steward Observatory, University of Arizona, 933 N Cherry Avenue, Tucson, AZ 85721, USA}

\author[0000-0003-4242-8606]{Ansh Gupta}
\affiliation{Steward Observatory, University of Arizona, 933 N Cherry Avenue, Tucson, AZ 85721, USA}

\author{Adam D. Myers}
\affiliation{Department of Physics \& Astronomy, University of Wyoming, 1000 E. University, Dept. 3905, Laramie, WY 82071, USA}

\author[0000-0003-3188-784X]{Nathalie Palanque-Delabrouille}
\affiliation{Lawrence Berkeley National Laboratory, 1 Cyclotron Road, Berkeley, CA 94720, USA}
\affiliation{IRFU, CEA, Universit\'{e} Paris-Saclay, F-91191 Gif-sur-Yvette, France}

\author[0000-0002-7633-431X]{Feige Wang}
\altaffiliation{NASA Hubble Fellow}
\affiliation{Steward Observatory, University of Arizona, 933 N Cherry Avenue, Tucson, AZ 85721, USA}

\author{Christophe Y{\`e}che} 
\affiliation{IRFU, CEA, Universit\'{e} Paris-Saclay, F-91191 Gif-sur-Yvette, France}

\author{Jessica Nicole Aguilar}
\affiliation{Lawrence Berkeley National Laboratory, 1 Cyclotron Road, Berkeley, CA 94720, USA}

\author[0000-0001-6098-7247]{Steven Ahlen}
\affiliation{Physics Dept., Boston University, 590 Commonwealth Avenue, Boston, MA 02215, USA}

\author{David M Alexander}
\affiliation{Centre for Extragalactic Astronomy, Department of Physics, Durham University, South Road, Durham, DH1 3LE, UK}

\author{David Brooks}
\affiliation{Department of Physics \& Astronomy, University College London, Gower Street, London, WC1E 6BT, UK}

\author{Kyle Dawson}
\affiliation{Department of Physics and Astronomy, The University of Utah, 115 South 1400 East, Salt Lake City, UT 84112, USA}

\author{Axel de la Macorra}
\affiliation{Instituto de F\'{\i}sica, Universidad Nacional Aut\'{o}noma de M\'{e}xico,  Cd. de M\'{e}xico  C.P. 04510,  M\'{e}xico}

\author[0000-0002-4928-4003]{Arjun Dey}
\affiliation{National Science Foundations's National Optical-Infrared Astronomy Research Laboratory, 950 N. Cherry Ave., Tucson, AZ 85719, USA}

\author[0000-0002-5402-1216]{Govinda Dhungana}
\affiliation{Department of Physics, Southern Methodist University, 3215 Daniel Avenue, Dallas, TX 75275, USA}

\author{Kevin Fanning}
\affiliation{Department of Physics, University of Michigan, Ann Arbor, MI 48109, USA}
\affiliation{The Ohio State University, Columbus, 43210 OH, USA}
\affiliation{University of Michigan, Ann Arbor, MI 48109, USA}

\author[0000-0002-3033-7312]{Andreu Font-Ribera}
\affiliation{Institut de F\'{i}sica d'Altes Energies (IFAE), The Barcelona Institute of Science and Technology, Campus UAB, 08193 Bellaterra Barcelona, Spain}

\author[0000-0003-3142-233X]{Satya Gontcho}
\affiliation{Lawrence Berkeley National Laboratory, 1 Cyclotron Road, Berkeley, CA 94720, USA}

\author{Julien Guy}
\affiliation{Lawrence Berkeley National Laboratory, 1 Cyclotron Road, Berkeley, CA 94720, USA}

\author{Klaus Honscheid}
\affiliation{Center for Cosmology and AstroParticle Physics, The Ohio State University, 191 West Woodruff Avenue, Columbus, OH 43210, USA}
\affiliation{Department of Physics, The Ohio State University, 191 West Woodruff Avenue, Columbus, OH 43210, USA}
\affiliation{The Ohio State University, Columbus, 43210 OH, USA}

\author{Stephanie Juneau}
\affiliation{National Science Foundations's National Optical-Infrared Astronomy Research Laboratory, 950 N. Cherry Ave., Tucson, AZ 85719, USA}

\author[0000-0003-3510-7134]{Theodore Kisner}
\affiliation{Lawrence Berkeley National Laboratory, 1 Cyclotron Road, Berkeley, CA 94720, USA}

\author[0000-0001-6356-7424]{Anthony Kremin}
\affiliation{Lawrence Berkeley National Laboratory, 1 Cyclotron Road, Berkeley, CA 94720, USA}

\author[0000-0001-7178-8868]{Laurent Le Guillou}
\affiliation{Sorbonne Universit\'{e}, CNRS/IN2P3, Laboratoire de Physique Nucl\'{e}aire et de Hautes Energies (LPNHE), FR-75005 Paris, France}

\author[0000-0003-1887-1018]{Michael Levi}
\affiliation{Lawrence Berkeley National Laboratory, 1 Cyclotron Road, Berkeley, CA 94720, USA}

\author{Christophe Magneville}
\affiliation{IRFU, CEA, Universit\'{e} Paris-Saclay, F-91191 Gif-sur-Yvette, France}

\author[0000-0002-4279-4182]{Paul Martini}
\affiliation{Center for Cosmology and AstroParticle Physics, The Ohio State University, 191 West Woodruff Avenue, Columbus, OH 43210, USA}
\affiliation{Department of Astronomy, The Ohio State University, 4055 McPherson Laboratory, 140 W 18th Avenue, Columbus, OH 43210, USA}
\affiliation{The Ohio State University, Columbus, 43210 OH, USA}

\author[0000-0002-1125-7384]{Aaron Meisner}
\affiliation{NSF's NOIRLab, 950 N. Cherry Ave., Tucson, AZ 85719, USA}

\author{Ramon Miquel}
\affiliation{Instituci\'{o} Catalana de Recerca i Estudis Avan\c{c}ats, Passeig de Llu\'{\i}s Companys, 23, 08010 Barcelona, Spain}
\affiliation{Institut de F\'{i}sica d'Altes Energies (IFAE), The Barcelona Institute of Science and Technology, Campus UAB, 08193 Bellaterra Barcelona, Spain}

\author[0000-0002-2733-4559]{John Moustakas}
\affiliation{Department of Physics and Astronomy, Siena College, 515 Loudon Road, Loudonville, NY 12211, USA}

\author[0000-0001-6590-8122]{Jundan Nie}
\affiliation{National Astronomical Observatories, Chinese Academy of Sciences, A20 Datun Rd., Chaoyang District, Beijing, 100012, P.R. China}

\author[0000-0002-0644-5727]{Will Percival}
\affiliation{Department of Physics and Astronomy, University of Waterloo, 200 University Ave W, Waterloo, ON N2L 3G1, Canada}
\affiliation{Perimeter Institute for Theoretical Physics, 31 Caroline St. North, Waterloo, ON N2L 2Y5, Canada}
\affiliation{Waterloo Centre for Astrophysics, University of Waterloo, 200 University Ave W, Waterloo, ON N2L 3G1, Canada}

\author{Claire Poppett}
\affiliation{Lawrence Berkeley National Laboratory, 1 Cyclotron Road, Berkeley, CA 94720, USA}
\affiliation{Space Sciences Laboratory, University of California, Berkeley, 7 Gauss Way, Berkeley, CA  94720, USA}
\affiliation{University of California, Berkeley, 110 Sproul Hall \#5800 Berkeley, CA 94720, USA}

\author[0000-0001-7145-8674]{Francisco Prada}
\affiliation{Instituto de Astrof\'{i}sica de Andaluc\'{i}a (CSIC), Glorieta de la Astronom\'{i}a, s/n, E-18008 Granada, Spain}

\author[0000-0002-3569-7421]{Edward Schlafly}
\affiliation{Space Telescope Science Institute, 3700 San Martin Drive, Baltimore, MD 21218, USA}

\author[0000-0003-1704-0781]{Gregory Tarl\'{e}}
\affiliation{University of Michigan, Ann Arbor, MI 48109, USA}

\author{Mariana Vargas Magana}
\affiliation{Instituto de F\'{\i}sica, Universidad Nacional Aut\'{o}noma de M\'{e}xico,  Cd. de M\'{e}xico  C.P. 04510,  M\'{e}xico}

\author{Benjamin Alan Weaver}
\affiliation{NSF's NOIRLab, 950 N. Cherry Ave., Tucson, AZ 85719, USA}

\author{Risa Wechsler}
\affiliation{Kavli Institute for Particle Astrophysics and Cosmology, Stanford University, Menlo Park, CA 94305, USA}
\affiliation{Physics Department, Stanford University, Stanford, CA 93405, USA}
\affiliation{SLAC National Accelerator Laboratory, Menlo Park, CA 94305, USA}

\author[0000-0001-5381-4372]{Rongpu Zhou}
\affiliation{Lawrence Berkeley National Laboratory, 1 Cyclotron Road, Berkeley, CA 94720, USA}

\author[0000-0002-4135-0977]{Zhimin Zhou}
\affiliation{National Astronomical Observatories, Chinese Academy of Sciences, A20 Datun Rd., Chaoyang District, Beijing, 100012, P.R. China}

\author[0000-0002-6684-3997]{Hu Zou}
\affiliation{National Astronomical Observatories, Chinese Academy of Sciences, A20 Datun Rd., Chaoyang District, Beijing, 100012, P.R. China}



\begin{abstract}
We report the first results of a high-redshift ($z \gtrsim 5$) quasar survey using the Dark Energy Spectroscopic Instrument (DESI). 
As a DESI secondary target program, this survey is designed to carry out a systematic search and investigation of quasars at $4.8<z<6.8$. 
The target selection is based on the DESI Legacy Imaging Surveys (the Legacy Surveys) DR9 photometry, combined with the Pan-STARRS1 data and $J$-band photometry from public surveys. 
A first quasar sample has been constructed from the DESI Survey Validation 3 (SV3) and first-year observations until May 2022. 
This sample includes more than 400 new quasars at redshift $4.7\le z<6.6$, down to 21.5 magnitude (AB) in the $z$ band, discovered from 35\% of the entire target sample. Remarkably, there are 220 new quasars identified at $z \ge 5$, more than one third of existing quasars previously published at this redshift. The observations so far result in an average success rate of 23\% at $z>4.7$. 
The current spectral dataset has already allowed analysis of interesting individual objects (e.g., quasars with damped Ly$\alpha$ absorbers and broad absorption line features), and statistical analysis will follow the survey's completion. 
A set of science projects will be carried out leveraging this program, including quasar luminosity function, quasar clustering, intergalactic medium, quasar spectral properties, intervening absorbers, and properties of early supermassive black holes.
Additionally, a sample of 38 new quasars at $z\sim 3.8-5.7$ discovered from a pilot survey in the DESI SV1 is also published in this paper. 
\end{abstract}

\keywords{galaxies: high redshift -- quasars: emission lines}


\section{Introduction} \label{sec:intro}
Quasars at $z \gtrsim 5$, residing in the cosmic times from the reionization era to the post-reionization epoch, are unique probes of the intergalactic medium (IGM) evolution during the last stage of the neutral hydrogen phase transition in the IGM \citep[e.g., ][]{eilers18, yang20b, bosman21} and the growth of early supermassive black holes (SMBH) \citep[e.g., ][]{shen19, yang21,farina22}. The study of quasar luminosity function and black hole mass function over cosmic epochs tracks SMBH accretions \citep[e.g., ][]{shen12, kelly13}. In addition, luminous high-redshift quasars are thought to trace the most massive dark matter halos in the young Universe \citep[e.g., ][]{costa14}. These luminous background sources also provide valuable sightlines for the investigations of intervening high-redshift H\,{\sc i} in the circumgalactic medium or interstellar medium, e.g., the damped Ly$\alpha$ (DLA) absorbers, and metal absorbers \citep[e.g., ][]{chen17, DOdorico18} . 

Recent successful high-redshift quasar surveys have increased the number of known $z\ge5$ quasars to $\sim$ 600 \citep[e.g.,][]{fan06,willott10,mortlock11,venemans15,wu15,banados16, jiang16, mazzucchelli17,mcgreer18, reed19,matsuoka19a,matsuoka19b,wang16,wang18,wang19,yang17,yang19a,yang19b,ross20,yang21} with the three most distant quasars at $z>7.5$ \citep{banados18,yang20a,wang21}. 
After the end of high-redshift quasar searches using the Sloan Digital Sky Survey I-IV \citep[e.g., ][]{schneider10, paris17, lyke20}, recent discoveries of high-redshift quasars are mainly based on single-object observations, which has a limited efficiency of spectroscopy. As a result, most existing high-redshift quasar surveys either focus on luminous quasars in a wide-field \citep[e.g.,][]{banados16,wang16,yang17} or target faint objects in a smaller area \cite[e.g.,][]{mcgreer18,matsuoka19a}.
The Dark Energy Spectroscopic Instrument \citep[DESI;][]{desi2022} employs a set of 10 multi-fiber spectrographs with wide spectral coverage, offering a unique opportunity to search for high-redshift quasars with high spectroscopic efficiency.  
The combination of deep wide-field imaging from the DESI Legacy Imaging Surveys \citep[][hereafter the Legacy Surveys]{dey19} and DESI multi-object spectroscopy will allow a new wide-field survey of high-redshift quasars, extending through the entire DESI footprint and to a fainter luminosity range.

In this paper, we report a DESI secondary program designed to search for quasars at $z \sim 4.8 - 6.8$ using DESI. This is our program's first publication to introduce the survey and publish discoveries from the DESI observations by 2022 May. We will describe the photometric datasets and target selection in Section 2. We then report the observations and the first sample of new quasars in Section 3. In Section 3, we will also present examples of scientific analyses using the current quasar spectra. In Section 4, in addition to a summary, we will discuss the future science projects using quasars from this survey. Additionally, we also publish a sample of new quasars from a pilot program of searching for $z \sim 4-5.3$ quasars during the DESI Survey Validation (SV) observations, which will be briefly described in Section 2. 
In this paper, we adopt a $\Lambda$CDM cosmology with parameters $\Omega_{\Lambda}$ = 0.7, $\Omega_{m}$ = 0.3, and H$_{0}$ = 70 $\rm km\,  s^{-1}\, Mpc^{-1}$. 
Photometric data in the optical are reported in the AB system after applying the Galactic extinction correction \citep{schlegel98, schlafly11}; photometric data from infrared surveys (e.g., in $J$, $W1$, and $W2$ bands) are on the Vega system. 

\section{A High-redshift Quasar Survey in DESI}
The DESI survey is designed to determine the nature of dark energy through the most precise measurement of the expansion history of the universe \citep{levi13}. 
With 5,000 fibers mounted in the focal plate and a 3-degree-diameter field of view \citep{desi16b,silber23}, DESI will measure the spectra of about 40 million galaxies and quasars covering 14,000 square degrees over its 5-year observing campaign \citep{desi16a,desi2022}. DESI target selection is based on the imaging data from the Legacy Surveys \citep[][Schlegel et al. 2023 in preparation]{zou17, dey19}, including targets of the Milky Way Survey \citep[MWS;][]{allende Prieto20, cooper23}, the Bright Galaxy Survey (BGS) \citep{ruiz-macias20, hahn23}, the Luminous Red Galaxy (LRG) sample \citep{zhou20, zhou23}, the Emission Line Galaxy (ELG) sample \citep{raichoor20,raichoor23}, and the quasar (QSO) sample \citep{yeche20, chaussidon23}. The details of the pipelines designed for DESI target selections, fiber assignments, survey operations, spectral reduction, and spectral classifications can be found in \cite{myers23}, Raichoor et al. (2023 in preparation), \cite{schlafly23}, \cite{guy23}, and Bailey et al. (2023 in preparation).
The DESI survey will not only use quasars as direct tracers of the matter distribution mostly at $z< 2.1$, but also use quasars at $z > 2.1$ as backlights for the intervening matter distribution via the Ly$\alpha$ forest. The quasar selection for the DESI main survey is based on a Random Forest algorithm and is mainly focused on quasars at $z\lesssim 4.5$ \citep{chaussidon23}.

We present an additional quasar survey program designed to systematically search for quasars at $z \gtrsim 5$ using DESI, which is part of the DESI secondary target program of the DESI survey. More details of DESI secondary target programs and the DESI target pipeline can be found in \cite{myers23}. 
The wide spectral coverage (3600 - 9800 \AA) and high efficiency of the DESI spectrographs allow sensitive identification of quasar Ly$\alpha$ lines up to redshift $\sim 6.8$ (covering wavelength range blueward of 1250 \AA\ in the rest-frame). In addition, DESI's spectral resolution (R $\sim 3000 - 5000$ at $\lambda >$ 5500 \AA) is higher than that of typical discovery spectra (R $\lesssim 1000$ ) used in most other high-redshift quasar surveys. Thus, the DESI spectra can be used to directly construct a dataset for a wide range of scientific analyses with no need of optical follow-up spectroscopy. For example, a complete DESI spectral dataset of $z \gtrsim 5$ quasars will allow us to investigate quasar luminosity function, quasar clustering, quasar rest-frame UV spectral properties, IGM evolution, and intervening absorbers, as well as the early supermassive black holes growth, together with multi-wavelength follow-up observations.

\subsection{Photometric Datasets}
The candidate selection of our survey is mainly based on the photometric data from the Legacy Surveys (\citealt{dey19}; Schlegel et al. 2023 in preparation), which images more than 19,700 square degrees of the extragalactic sky visible from the Northern hemisphere in $g, r$, and $z$ bands. The Legacy Surveys consists of three programs: the Beijing-Arizona Sky Survey \citep[BASS;][]{zou17}, covering an area in the North Galactic Cap with Dec $> 32.375$ deg in $g$ and $r$ bands, the Mayall $z$-band Legacy Survey \citep[MzLS;][]{dey19}, imaging the same area as BASS in $z$ band, and the Dark Energy Camera Legacy Survey \citep[DECaLS;][]{dey19}, mapping the entire South Galactic Cap and the regions in North Galactic Cap with Dec $< 34$ deg in $g$, $r$, and $z$ bands. There are also other public DECam $grz$ data within the DESI footprint. For example, the data from the Dark Energy Survey \citep[DES;][]{des05}, which overlaps with the Legacy Surveys in an era of 1,130 deg$^{2}$. The Legacy Surveys program makes use of the DES raw data instead of re-observing that era.

The Legacy Surveys DR9\footnote{https://www.legacysurvey.org/dr9/} includes images in all the three bands with $\ge 3$ observational passes over 14,750 square degrees, with median PSF depth of 24.7 (24.2 in BASS area), 24.2 (23.7 in BASS), and 23.3 in the $g$, $r$, and $z$ bands, respectively. 
The optical data are complemented by photometry in infrared bands ($W1$ and $W2$) from the all-sky data of the Wide-field Infrared Survey Explorer \citep[{\it WISE};][]{wright10}. The $W1$ and $W2$ data in the Legacy Surveys catalog are based on forced photometry in the {\it unWISE} images at the locations of the Legacy Surveys optical sources.
The {\it unWISE} images are the unblurred coadds of {\it WISE} \citep{meisner17,meisner18} including all imaging through year 6 of NEOWISE-Reactivation \citep{mainzer14}.  

We have also included photometric data from the Pan-STARRS1 survey \citep[PS1;][]{chambers16}, which covers the $3\pi$ sky at $Dec > -30$ deg in the optical $grizy$ bands. In our survey, the PS1 $i$-band photometry is required for target selection, and data in PS1 $z$ and $y$ bands are included if available (see details in the next section). The PS1 survey has 5$\sigma$ depth of 23.1, 22.3, and 21.4 magnitude in the $i$, $z$, and $y$ bands, respectively. 
In the following discussions of color selection, we use $z$ to represent the $z$-band photometric data from the Legacy Surveys and use $z_{\rm P1}$ for the PS1 photometry. 

In addition, in order to further reduce contamination rate of quasar selection, photometric data in the near-infrared (NIR) $J$-band are used for objects with available $J$-band photometry. We collect $J$-band data from the three public large-area NIR imaging surveys: the UKIRT Hemisphere Survey (UHS) \citep{dye18}, the UKIRT InfraRed Deep Sky Surveys--Large Area Survey \citep[ULAS,][]{lawrence07}, and the VISTA Hemisphere Survey \citep[VHS,][]{mcmahon13}. The UHS and ULAS surveys cover an area of 17,900 square degrees in the northern sky with a depth of 19.6 mag (Vega) in $J$. The VHS survey aims to map the entire Southern sky and has a depth of 20.2 mag (Vega) in the $J$ band.

\subsection{DESI Survey of Quasars at $z \sim 4.8 - 6.8$} 
The quasar selection of our survey is based on optical-infrared color cuts, which have been previously applied to successful $z\sim5-6.5$ quasar surveys \citep[e.g.,][]{banados16,wang16,wang19,yang17,yang19a,yang19b}.  
We start our selection with the Legacy Surveys DR9 catalog and first apply a set of standard cuts: 1) {\tt BRICK\_PRIMARY} = T to select objects within the brick boundary; 2) {\tt MASKBITS} not in [1, 10, 12, 13], which requires that the targets do not touch pixels in the vicinity of bright stars, ``bailout" blob (areas leading to fitting issue due to a very high source density), large galaxies, or globular clusters; 3) {\tt TYPE} = PSF OR ({\tt dchisq}[1] -- {\tt dchisq}[0])/{\tt dchisq}[0] $<$ 0.01 to select objects with stellar morphology. 
We next require that all targets should be observed in all bands and limit the signal-to-noise (S/N) in the $z$ ($> 5\sigma$), $W1$ ($> 3\sigma$), and $W2$ ($> 2\sigma$) bands. Then we apply very relaxed cuts in the $gr$ bands (i.e., $S/N_{g} < 5$ or $g > 24.5 $ or $g-r > 1.8$; $S/N_{r}< 5$ or $r-z>1.0$) to reduce the sample size. These generate a pre-selected sample with $\sim$ 13 million sources within the DESI footprint for following color-color selection. 

\begin{figure}%
\centering 
\epsscale{1.17}
\plotone{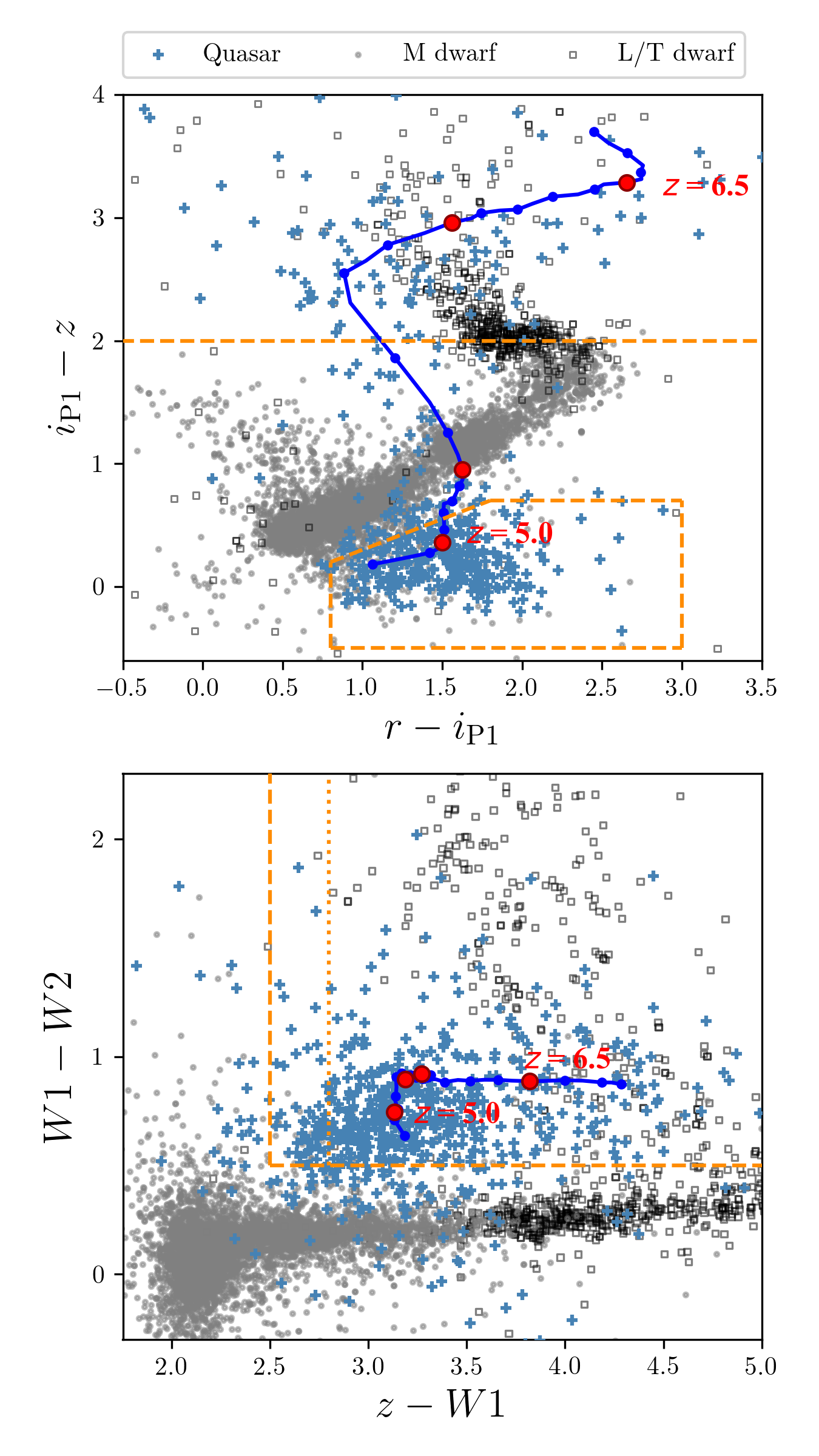} 
\vspace{-35pt}
\caption{{\bf Top:} The $riz$ color-color diagram used for the quasar selection. The blue solid line shows the color track of quasars from $z=4.8$ to $z=6.8$ with a step of 0.1 (blue filled circles). The red points on the track represent quasars at $z = 5.0, 5.5, 6.0$, and 6.5. The blue crosses denote all existing quasars known at $4.8 \le z \le 6.8$ \citep[e.g.,][]{fan06,willott10,mortlock11,venemans15,wu15,banados16,banados18, jiang16, mazzucchelli17,mcgreer18,matsuoka19a,matsuoka19b,paris17,reed19,wang16,wang18,wang19,wang21,yang17,yang19a,yang19b,yang20a,yang21}, while the grey points and the black squares show the loci of M and L/T dwarfs \citep{kirkpatrick11,mace14,best15}, respectively. The orange dashed lines bound the selected regions for $z \sim 5$ and $z \sim 6$ quasar candidates. {\bf Bottom:} The $zW1W2$ color-color diagram. All symbols are the same as in the top panel. The orange dashed lines represent the cuts used for $z \sim 5$ and $z \sim 6$ quasar candidates. A more stringent $z-W1$ cut (orange dotted line) is used for the selection of $z > 6.4$ quasars. Note that the photometry in $W1$ and $W2$ are in the Vega system ($W1_{\rm Vega} = W1_{\rm AB} - 2.699$, $W2_{\rm Vega} = W2_{\rm AB} - 3.339$).}
\label{fig:color1}
\end{figure}

One basic technique used for the color selection of high-redshift quasars is based on color `drop-out', which is caused by the Ly$\alpha$ break in quasar spectra due to significant IGM absorption at the wavelength blueward of the Ly$\alpha$ emission line at these redshifts. The Ly$\alpha$ line of a $z \sim 5$ quasar is located in the $i$ band (7294.02 $\AA$ at $z=5$). Thus, the $i$ band is typically used as a detection band for the selection of $z \sim 5$ quasars, and the $g$ and $r$ are the dropout bands. At $z \gtrsim 5.7$, Ly$\alpha$ line moves into the $z$ band and the $i$ band becomes the dropout band, so the $i$ band also plays a critical role in the selection of $z \sim 6$ quasars. 
We therefore include the $i$-band photometric data from the PS1 survey. 

We cross match the pre-selected sample selected above with the PS1 DR1 and DR2 catalogs. We only use DR1 data when an object is not in the DR2 catalog but is in the DR1. 
Then we build up different selection criteria for three redshift ranges, $z \sim 4.8-5.4$, 5.7 -- 6.4, and 6.4--6.8, according to the quasar color tracks in different color-color spaces. 
The main color-color selections are the $r-i/i-z$ for $z \sim5 $ and 6, and the $z-y/y-W1$ for $z \sim 6$ and 6.5. The $z-W1/W1-W2$ color cuts are used for all redshifts. 
The redshift range of $z \sim 5.4 -5.6$ is the redshift at which quasar locations fully overlap with those of middle-type M dwarfs in the $riz$ color-color space. Multiple NIR colors are necessary to select quasars in this redshift range, as demonstrated in our previous successful $z\sim 5.5$ quasar survey \citep{yang17,yang19a}. Thus we exclude this redshift range in this survey. Below we briefly describe our selections of quasars in the three redshift ranges, and show the main color-color diagrams in Figures \ref{fig:color1} and \ref{fig:color2}. The selection criteria are listed in Appendix \ref{A}.

{\it $z \sim {\it 4.8-5.4}$ quasar candidates} -- {\bf 1)} We first select objects with $g$-band dropout by applying the cuts of S/N$_{g} < 3$ or $g-r > 2.5$. {\bf 2)} We next require objects to have $> 5 \sigma$ detection in the $i_{\rm P1}$ band and apply the $r-i/i-z$ color cuts as shown in Figure \ref{fig:color1}. {\bf 3)} Then we apply the $z-W1/W1-W2$ cuts (Fig. 1, bottom). {\bf 4)} Additionally, for objects that have $> 3 \sigma$ detections in the $y_{\rm P1}$ band, we use a $y_{\rm P1}-W1$ color (Fig. \ref{fig:color2}) to improve the purity. All objects without $y_{\rm P1}$ data are kept.

{\it $z \sim {\it 5.7-6.4}$ quasar candidates} -- 
At $z \gtrsim 6 $, the Ly$\alpha$ emission line moves into the $z$ band.  {\bf 1)} Thus we first require the objects to be dropouts in the $g$ and $r$ bands (see criteria in Appendix A). {\bf 2)} Then we select objects without $i$-band detection (S/N$_{i\rm P1} < 3$) or meeting the cut of $i_{\rm P1}-z > 2.0$. {\bf 3)} We also apply the $z-W1/W1-W2$ cuts, which are the same to the $z \sim 5$ selection. {\bf 4)} In addition, if one object has $>3 \sigma$ detections in the PS1 $z$ and $y$ bands, $z_{\rm P1}$ and $y_{\rm P1}$ photometry are employed to further reduce the contamination rate. The $z_{\rm P1}$ band covers a wavelength range bluer than the Legacy Surveys' $z$ band, so it can provide additional constraints on colors. We limit $y_{\rm P1}-W1 >$ 2.399 and $z_{\rm P1}-y_{\rm P1} < 1.0$, as shown in Figure \ref{fig:color2}. Objects without $z_{\rm P1}$ and $y_{\rm P1}$ detections are kept.

{\it $z > {\it 6.4}$ quasar candidates} -- {\bf 1)} We first select objects that are dropouts in the $gri$ bands (see criteria in Appendix \ref{A}).  {\bf 2)} At the same time, we apply a more stringent $z-W1$ cut as shown in Figure \ref{fig:color1} (bottom). {\bf 3)} At such high redshift, most quasars only have the $z$-band detection, except for very luminous ones, which could be detected in the $i$ band due to a bright Ly$\beta$ line. Thus, we require additional photometric data in the $y_{\rm P1}$ band for all candidates. We select objects with $> 3 \sigma$ detection in the $y_{\rm P1}$ and require $y_{\rm P1}-W1 >$ 2.399 (Fig. \ref{fig:color2}).  {\bf 4)} We reject objects that have $>3 \sigma$ detections in $z_{\rm P1}$ band but have colors of $z_{\rm P1}-y_{\rm P1} \le1.0$ (Fig. \ref{fig:color2}) or $z_{\rm P1} - W1< 4.0$.

\begin{figure}%
\centering 
\epsscale{1.2}
\plotone{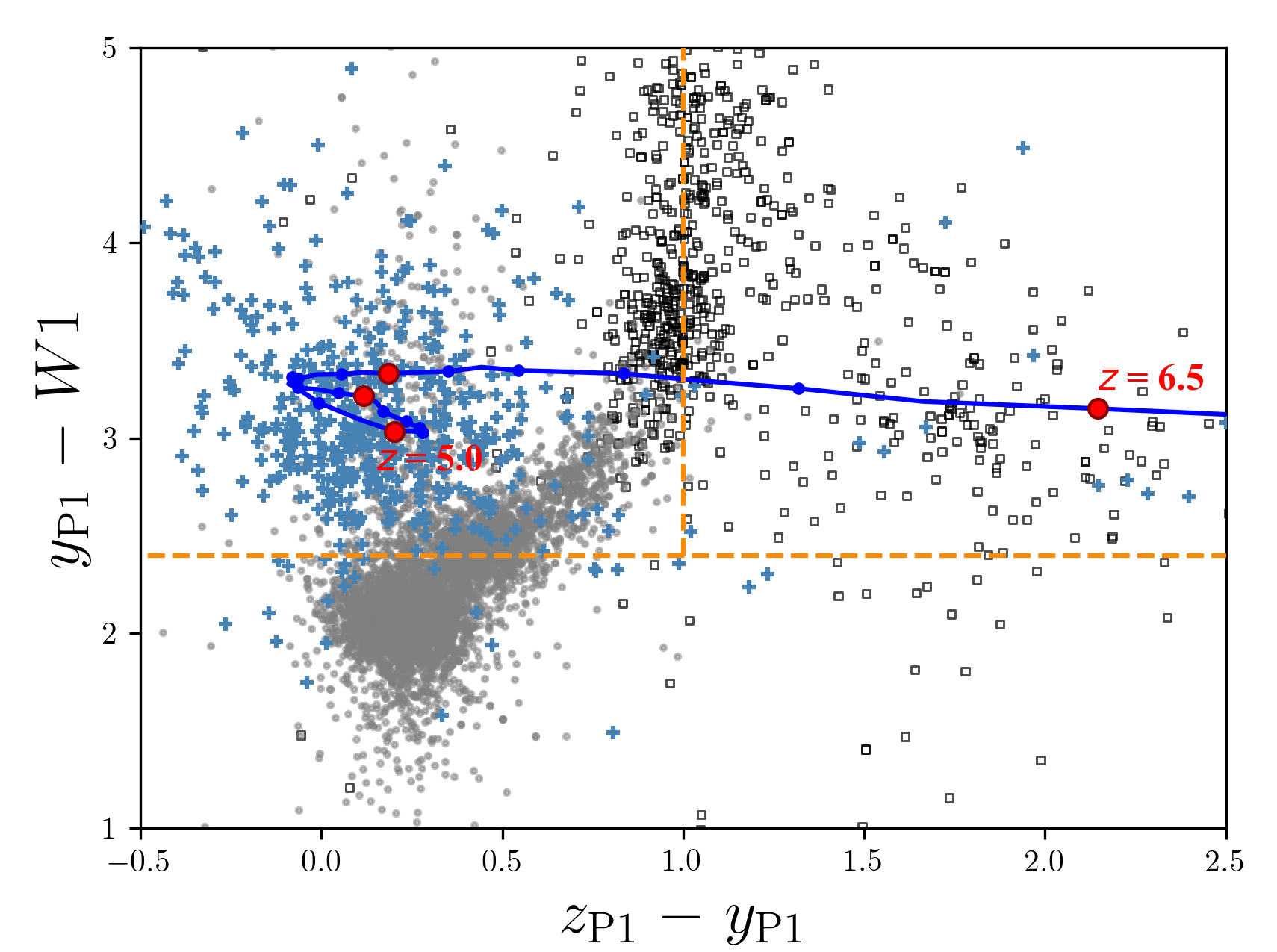} 
\vspace{-15pt}
\caption{The $z_{\rm P1}y_{\rm P1}W1$ color-color diagram, used for the selection of $z \sim 6$ and 6.5 quasars. All symbols are the same as Figure \ref{fig:color1}. A $y_{\rm P1} - W1$ cut is applied for all objects, and the two regions separated by the $z_{\rm P1}-y_{\rm P1}$ color (vertical orange dashed line) are the selections for $z \sim 6$ and $z > 6.4$ quasar candidates, respectively. }
\label{fig:color2}
\end{figure}

NIR photometry can play an important role in further rejecting M/L/T dwarf contaminants in high-redshift quasar selections by leveraging the power-law continuum of quasars. Among the NIR broad bands (i.e., $YJHK$), the $J$ band has been mapped in the widest sky coverage. We therefore include $J$-band data into our selection, using photometry from the UHS, ULAS, and VHS surveys. We reject all objects that are covered by $J$-band photometry but have bad $J$-based colors (e.g., $J-W1 < 1.5$ or $y_{\rm P1} - J > 2.0$ if $y_{\rm P1}$ is available; $J_{\rm AB}$ = $J_{\rm Vega} + 0.938$). Objects without $J$-band data are kept. 

In this first year of DESI observations, our high-redshift quasar selection focuses on a relatively bright candidate sample, with a $z$-band magnitude limit of 21.4 for $z\sim5$ quasar candidates and 21.5 for $z \sim 6 $ as well as $z > 6.4$ candidates. This is already one magnitude deeper than the previous wide-area $z \sim 5$ quasar surveys \citep{wang16,yang17,yang19a}, while this depth only covers the luminous end of $z > 6$ quasar population. We will extend the quasar selections towards a fainter range in subsequent years of the DESI survey.

In total, we obtain $\sim$ 3,500 targets for $z\sim5$ quasar selection, $\sim$ 3,000 targets from $z\sim$ 6 selection, and $\sim$ 450 targets for $z > 6.4$ quasars within the DESI footprint, with an average target density of 0.5 per square degree. These targets have a {\tt bit-name} of Z5\_QSO in the DESI target catalog. We are not excluding known quasars that pass our selection criteria in order to construct a uniformly observed spectral dataset, which is important for future spectral analysis. Multi-epoch spectra will also allow the study of quasar variability. 
At the survey depth, about 60-90\% of known quasars can be selected by our selections in three different redshift ranges, as shown in Figure \ref{fig:hist}, which can be treated as a rough estimate of the selection completeness. A careful measurement of the selection function using a simulated quasar sample will be presented with our final quasar sample.

\subsection{Selection of Quasars at $z \sim 4 - 5.3$ in SV1}
In this paper, we also report results from a selection of $z\sim 4 - 5.3$ quasars using photometric data from the Legacy Surveys only, and this selection is used as a pilot selection during the DESI SV1 \citep{myers23, desi23a}. DESI conducted SV observations, prior to the main, five-year mission, to validate the systems, to refine the purity and completeness of the targeting algorithms, and to stress-test the procedures that would be needed \citep{desi23a}. SV1 is the first iteration of SV ran. The pilot survey of $z \sim 4 - 5.3$ quasars was designed to test the quasar selection without the $i$-band photometry. The selection was based on $grz$ and $zW1W2$ colors. The selection first adopted the $g$-band and $r$-band dropout techniques. The $r-z$ color was applied instead of the $riz$ colors, and the $zW1W2$ cuts were used with a relaxed $W1-W2$ color ($W1-W2 > 0.3$) for $z < 4.8$ quasars due to the bluer $W1-W2$ color of $z \sim 4.5 $ quasars. The selection criteria are listed in the Appendix \ref{B}. 

This selection yielded $\sim$ 60 quasars during the SV1 observations between 2020 December and 2021 May. Among them, there were 38 new quasars at $3.8 \le z \le 5.7$, as listed in Table \ref{tab:sv1quasar} and shown in Figure \ref{fig:sv1spectra} in the Appendix \ref{B}. 
The selection of $z\sim 5$ quasars without $i$-band photometry could help to construct a quasar sample that is independent of the Ly$\alpha$ line luminosity. Indeed, we were able to discover weak line (WL) quasars and broad absorption line (BAL) quasars at $z \sim 5$ (Fig. \ref{fig:sv1spectra}), which are missed by the selection mentioned above and also by the quasar surveys in previous works \citep[e.g.,][]{wang16,yang17,yang19a}. 
However, the lack of $i$-band photometry led to a high contamination rate;  the success rate of this selection was only 2-3\%, which is quite low compared to the efficiency of our main selection described above (see details of efficiency in Section 3). Therefore, this selection was not retained after the SV observations. We publish the 38 new quasars discovered in SV1 in this paper.

\section{Results from Observations in the First Year}
\subsection{Observations and Redshift Measurements}
The selected $z \sim 4.8 - 6.8$ quasar candidates are mainly observed as dark-time targets in the DESI Main Survey, which started on May 14, 2021. A minor part (1.6\%) of candidates were observed during the DESI 1\% survey (SV3; in 2021 April) and the DESI SV1 (before April 4, 2021; for 27 targets overlapped with the SV1 selection). These targets have the same fiber assignment priority to the primary QSO targets \citep{schlafly23}.
All our targets are designed to first have a single-exposure with an effective exposure time\footnote{The effective exposure time is defined based on the spectroscopic average signal-to-noise ratio and is utilized to achieve required S/N in a minimum amount of time. More details can be found in \cite{guy23} and \cite{schlafly23}. The exposure times reported later in this paper are all the actual exposure times.} \citep{guy23} of 1000s for the purpose of identification. A S/N $>$ 3 per pixel is expected on the Ly$\alpha$ line to identify a target. Then all targets identified as quasars will be observed with three repeat exposures by DESI for high quality spectra.
As of May 14, 2022, there are $\sim$ 2370 targets that have been observed from our program, which is about 35\% of the entire candidate sample. The spectra of our targets after a first-pass observation have sufficient S/N for quasar identification, as shown in Figure \ref{fig:Specexample} and Figure \ref{fig:allspectra}.

We identify quasars using the daily coadd spectra in the database, which are products of the DESI spectroscopic pipeline \citep{guy23}. The spectroscopic pipeline is designed to extract spectra from the raw data, subtract sky model, flux-calibrate spectra based on standard star exposures, and then measure their classifications as well as redshifts. We identify quasars via visual inspection and estimate quasar redshifts from visual spectral fitting, which is not included in the general DESI spectra visual inspection \citep{alexander23, lan23}.
The visual spectral fitting is based on a semi-automated toolkit, A Spectrum Eye Recognition Assistant \citep[ASERA;][]{yuan13}, which has been used to measure redshifts for a number of high-redshift quasars \citep[e.g.,][]{wang16,wang19,yang17,yang19a}. All spectra of our program have been visually inspected by at least two inspectors. 
When fitting spectra, we match the observed spectrum with the SDSS quasar template based on quasars' broad emission lines, Ly$\beta$, Ly$\alpha$, O\,{\sc i}, Si\,{\sc iv}, and C\,{\sc iv} if visible, as well as the continuum emission. Figure \ref{fig:Specexample} shows three examples of the DESI spectra with emission lines identified. They represent the quality of single-epoch observation spectra used for our identification and visual spectral fitting. Such redshift measurements have a typical uncertainty of $\pm 0.03$ \citep[e.g.,][]{wang16,yang17}. For weak line quasars and strong broad absorption line quasars, the uncertainty could be as large as $\sim$ 0.05 - 0.1. We have one primary inspector and at least one secondary inspector. If redshifts from primary and secondary inspectors have a difference $\le 0.03$, we use redshift from the primary inspector. For a few cases that we get a discrepancy $> 0.03$, the inspectors repeat the VI together until they obtain a consistent result.

\begin{figure}%
\centering 
\epsscale{1.26}
\plotone{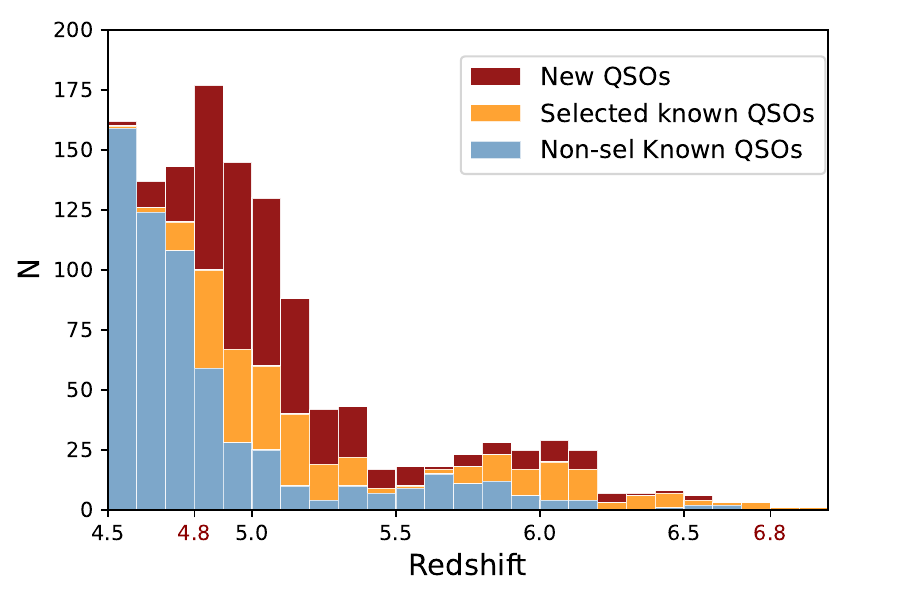} 
\vspace{-15pt}
\caption{A stacked histogram presenting the redshift distribution of new quasars (red, ``New QSOs") from this survey, compared with all existing known quasars that are in the DESI footprint and within our survey depth (i.e., $<$ 21.5 in the LS $z$ band). Known quasars that pass our selections are shown in orange (``Selected known QSOs"), while the quasars that do not meet our selections are in the color of blue (``Non-sel Known QSOs"). The redshift range that our survey focuses on has been marked in red ($z = 4.8 - 6.8$).}
\label{fig:hist}
\end{figure}

\begin{figure}%
\centering 
\epsscale{1.2}
\plotone{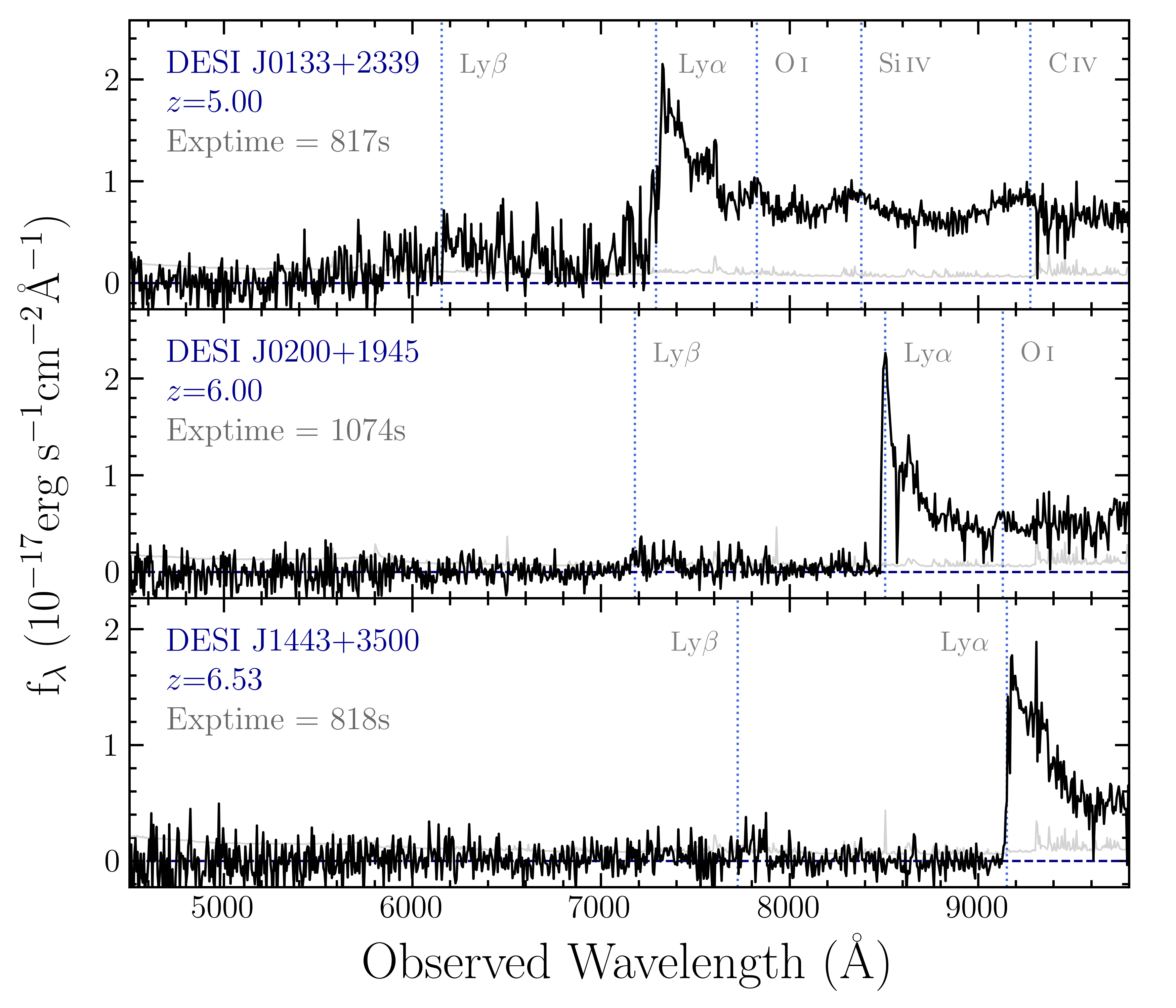} 
\vspace{-15pt}
\caption{Examples of the DESI spectra for quasars at $z = 5$, $6$, and $6.5$, binned with seven pixels. The three spectra are from observations with $\sim 800 - 1000$s actual exposure time. These three quasars are 20.55, 20.72, and 21.20 magnitude in the Legacy Survey $z$ band. The quasar broad emission lines (i.e., Ly$\beta$, Ly$\alpha$, O\,{\sc i}, Si\,{\sc iv}, and C\,{\sc iv}) used for redshift measurements are marked with blue dotted lines. The spectra of all 412 new quasars are shown in Figure \ref{fig:allspectra}.}
\label{fig:Specexample}
\end{figure}

\subsection{The Discovery of More than 400 New Quasars}
Among the observed targets, there are 556 quasars identified in the redshift range of $z \sim 4.4 - 6.8$, including 144 known quasars and 412 new quasars. These new quasars span a redshift range from 4.44 to 6.53, with 220 quasars at $z \ge 5$ and 25 quasars at $z \ge 6$. The redshift distribution of these quasars is presented in Figure \ref{fig:hist}. So far, the observations of these targets result in an average success rate of 23\% of quasars at $z>4.7$. If we count the success rate of $z \sim 4.8-5.4$, $z \sim 5.7 - 6.4$, and $z > 6.4$ quasar selections separately, we obtain 39\%, 8\%, and 5\%, respectively. The main contaminants are M/L/T dwarfs ($>$ 90\%) and red galaxies, as shown in the Appendix \ref{C} (Fig. \ref{fig:contam}). In Figure \ref{fig:Specexample}, we show examples of $z \sim$ 5, 6, and 6.5 quasars newly identified from our program. Quasar DESI J144355+350055 at $z=6.53$ is the highest-redshift new quasar identified from our survey to date. 

Although our survey has not been completed, it has already significantly increased the number of quasars known at $z \ge 4.8$ (377 new quasars). The total number of known quasars at $z= 4.8 -6.8$ from previous works is about 810, and thus the new quasars from this program already expand the existing known quasar sample by 46\%. At $z \ge 5$, our survey so far has increased the known quasar sample by more than one third, with 220 new discoveries and 628 existing known quasars.
Within the DESI footprint, our new discoveries have more than doubled the quasar sample size in the redshift range of $4.8 -5.4$. 

We list all 412 new quasars in Table \ref{tab:newquasar} and plot all spectra of these new quasars in Fig \ref{fig:allspectra}. The spectra plotted here are based on the coadd data of the DESI daily spectra. All spectral data will be made public in digital form in future DESI data releases \citep[e.g.,][]{desi23b}.

\subsection{Examples of Scientific Analysis}
So far, the DESI high-redshift survey and the repeat observations of newly identified quasars have not been finished, so this sample is not complete enough for statistical analysis. However, there are already a number of spectra that have sufficient quality to allow us to start science analysis for individual objects.

\begin{figure}%
\centering 
\epsscale{1.2}
\plotone{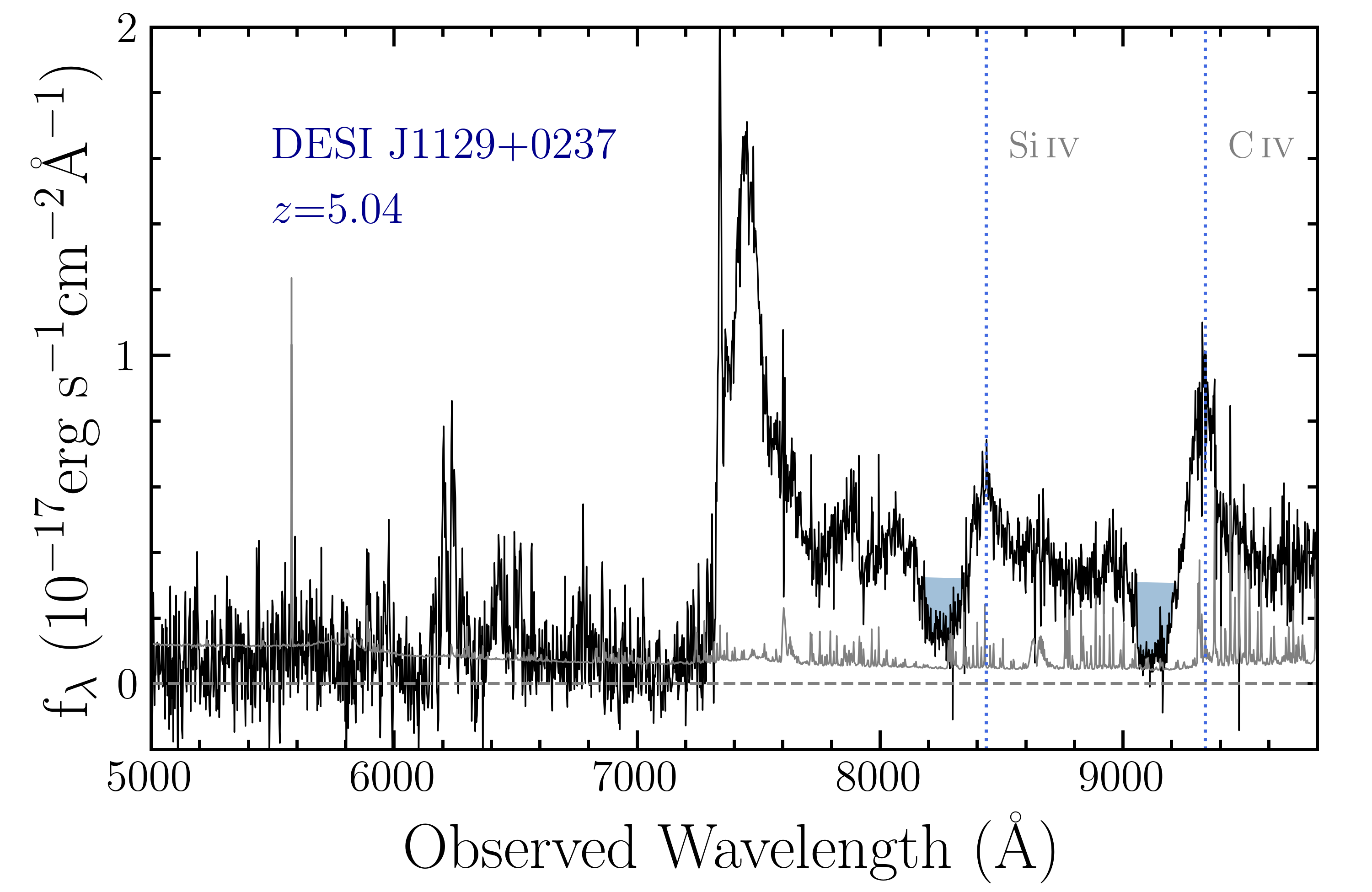} 
\vspace{-15pt}
\caption{The DESI spectrum of a $z=5.04$ quasar J112903.54+023706.9, coadded from four exposures with a total exposure time of $\sim$ 4,400s. The spectrum plotted here is binned with 3 pixels for the BAL absorption measurements. The grey line is the spectral uncertainty. The target has a $z$-band magnitude of 21.3. This spectrum has sufficient S/N for quasar classification and study of strong absorption features. The Si\,{\sc iv} and C\,{\sc iv} BAL absorption troughs are denoted by the blue shaded regions.}
\label{fig:j1129}
\end{figure}

As an example, Figure \ref{fig:j1129} shows the spectrum of quasar DESI J112903.54+023706.9, which has a $z$-band magnitude of 21.3, among the faintest targets in our candidate sample. The spectrum has $\sim$ 4,400s exposure time, and the average S/N over the Ly$\alpha$+N\,{\sc v} line region is 9 per 0.8 \AA\ pixel and 3 per pixel on continuum. All quasars from this survey will be observed repeatedly by DESI to have comparable or longer exposure time. 
In addition, as shown in Figure \ref{fig:j1129}, the Si\,{\sc iv} and C\,{\sc iv} BAL features can be seen clearly from this spectrum. The balnicity index BI \citep{weymann91} is derived as $\sim$ 2000 km s$^{-1}$ and 2500 km s$^{-1}$ for the Si\,{\sc iv} and C\,{\sc iv} absorptions, respectively, using a 3-pixel binned spectrum. 
The BI \citep{weymann91} is calculated by 
${\rm BI} = \int_{v_{\rm min}}^{v_{\rm max}}\left ( 1 - \frac{f(v)}{0.9} \right )C dv$,
where $f(v)$ is the normalized spectrum, and $C$ is set to 1 only when $f(v)$ is continuously smaller than 0.9 for more than 2000 km s$^{-1}$, otherwise it is set to 0.0. The value of $v_{\rm min}$ is set to 0.
The final spectral dataset will allow us to identify BAL quasars and measure the BAL fractions at different redshift bins.

\begin{figure}%
\centering 
\epsscale{1.2}
\plotone{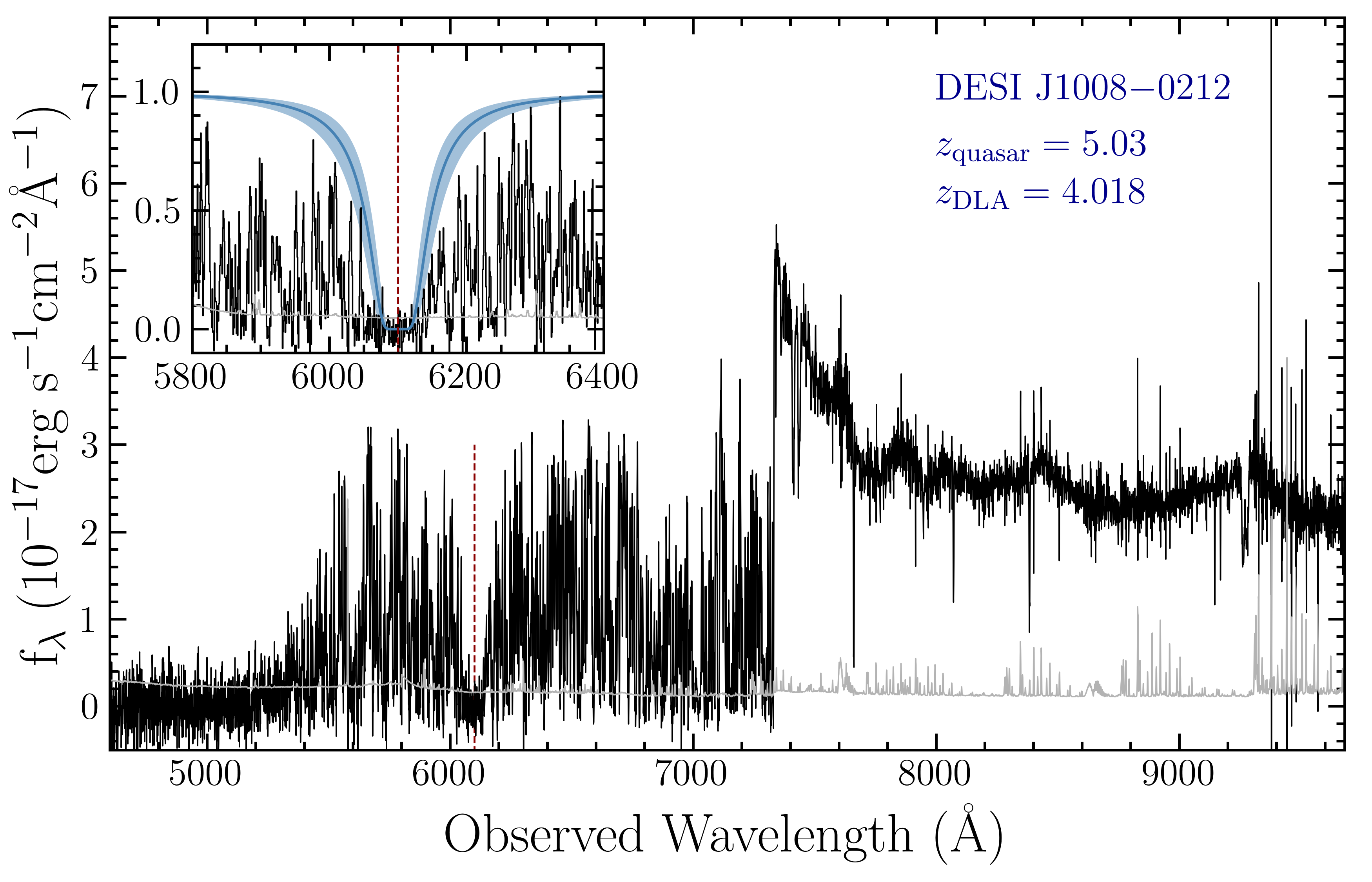} 
\vspace{-15pt}
\caption{The DESI spectrum of a $z=5.03$ quasar J100828.30$-$021229.8 without binning. The grey line represents the spectral uncertainty. A DLA system at $z=4.018$ (red dashed line) is clearly present in the absorption spectrum. Its redshift is estimated using metal lines. The Voigt profile fitting (inset plot) yields a column density of ${{\rm log}~N_{\rm HI}}=21.2\pm0.2$.}
\label{fig:dla}
\end{figure}

These quasar spectra also enable us to search for high-redshift DLA absorbers. Figure \ref{fig:dla} shows an example of a $z = 4.018$ DLA identified in the spectrum of $z = 5.03$ quasar DESI J100828.30$-$021229.8. The spectrum has a coadd exposure time of $\sim$ 5,000s. The redshift of this DLA is derived using a set of metal lines. Metal lines \ion{Si}{2} $\rm \lambda 1526$, \ion{C}{4}  $\rm \lambda 1548$, \ion{C}{4}  $\rm \lambda 1550$, \ion{Fe}{2} $\rm \lambda 1608$, and \ion{Al}{2} $\rm \lambda 1670$ have been identified. The column density of this DLA is determined as ${{\rm log}~N_{\rm HI}}=21.2\pm0.2$ by fitting a Voigt profile.
We place the centroid of a Voigt profile to the redshift of the low-ion metal-line transitions ($z = 4.018$) and manually select values of $N_{\rm HI}$ to fit the profile, assuming a single component of DLA plus possible blended Ly$\alpha$ forest in the wings. The uncertainties of $N_{\rm HI}$ estimations for high-redshift DLAs are dominated by continuum uncertainty and Ly$\alpha$ forest line blending \citep{rafelski12}. 
The column density of metal lines is derived from the apparent optical depth method using the routines publicly available in the LINETOOLS package \citep{prochaska17}. 
We then obtain log$N = 14.64 \pm 0.03$ for Si using the \ion{Si}{2} $\lambda$1526 line, which implies a metallicity of $-2.06\pm0.21$. This follows the DLA metallicity-redshift evolution trend \citep{rafelski12}. 
A systematical search will be carried out using the complete DESI spectral sample. 

In addition, we study radio detections among these new quasars by cross-matching (2$''$) new quasars with the catalogs from the Faint Images of the Radio Sky at Twenty-cm \citep[FIRST;][]{becker95}, the first epoch of the Very Large Array Sky Survey \citep[VLASS;][]{lacy20, gordon20}, and the second data release from the ongoing LOw-Frequency ARray (LOFAR) Two-metre Sky Survey (LoTSS) \citep{shimwell22}. The cross-match radius is chosen following recent LOFAR studies \citep[e.g.,][]{retana-montenegro18, gloudemans22}, although it might be small for matching FIRST catalog. As a first test, we choose this radius for a reliable detection. We find six quasars detected in the FIRST catalog with peak flux densities from 1.2 to 54.7 mJy/beam at 1.4 GHz. The same six objects are in the VLASS catalog. Two quasars have significant differences in the peak flux density between the measurements from FIRST and VLASS, which might be useful to study quasar radio variability. 
We also obtain eleven additional radio detections ($z \sim 4.7 - 6.0$) in the LOTSS DR2 catalog, with peak flux densities from 0.3 to 1.1 mJy/beam at 144 MHz. 
In the future, using the final sample, we will be able to measure the radio-loud fractions in different redshift bins.

\section{Summary and Future Studies}
In this paper, we describe a high-redshift ($z \sim$ 4.8--6.8) quasar survey with DESI, as a secondary program of the DESI survey. Using the photometric data from the Legacy Surveys DR9 ($g,r,z,W1$, and $W2$), the PS1 survey ($i_{\rm P1}$, $z_{\rm P1}$, and $y_{\rm P1}$), and several NIR surveys ($J$), we carried out selections of quasars in three redshift ranges: $z \sim 4.8-5.4$, $5.7 - 6.4$, and $6.4-6.8$. The observations during the DESI SV1 and the main survey before May 14, 2022 have covered 35\% of candidates and identified more than 550 quasars, yielding 412 new quasars at $4.44 \le z \le 6.53$, with 377 quasars at $z \ge 4.8$ and 220 at $z \ge 5.0$. The observations to date result in an average success rate of 23\%. Assuming this success rate, we would expect $\sim$ 1,000 new quasars from this survey. The new discoveries from this survey have increased the sample size of known quasars by $\sim$ 46\% in the redshift range that we are targeting. The high quality spectra allow us to start the study of individual quasar spectra, although the current dataset is not yet complete enough for statistical analysis.

Based on this survey, we expect to construct a large $z \gtrsim 5$ quasar sample for scientific analyses.
We will present details of the full quasar sample and the selection function when the identification of the entire sample is complete. The quasar luminosity functions will be also calculated at that time. In subsequent papers, we will present spectral fitting and basic quasar spectral properties (e.g., continuum slope, luminosity, and line width).
When all repeat observations of quasars have been completed, the DESI quasar spectral dataset will allow a set of science projects, including but not limited to $z\sim 5-6$ quasar clustering, quasar proximity zone, IGM evolution, radio-loud fraction, WL and BAL quasar fractions, as well as intervening DLA and metal absorbers. Follow-up observations in the NIR will help to build a sample for the study of early SMBHs growth and the BH mass function. We will also push the high-redshift quasar survey to a fainter magnitude in the future DESI survey.

\startlongtable
\begin{deluxetable}{l l l c c}
\tablecaption{The 412 New Quasars from Our Main Selection.}
\tabletypesize{\footnotesize}
\tablewidth{\textwidth}
\tablehead{
\colhead{Name} &
\colhead{RA} &
\colhead{Dec} &
\colhead{$z$\tablenotemark{a}} &
\colhead{$z_{\rm LS}$} 
}
\startdata
  DESI J000147.64--035247.4 & 0.44850 & --3.87986 & 5.27 & 20.79$\pm$0.01\\
  DESI J000232.92+131433.7 & 0.63716 & 13.24270 & 5.82 & 21.30$\pm$0.04\\
  DESI J000233.24+212725.0 & 0.63850 & 21.45695 & 6.19 & 21.47$\pm$0.04\\
  DESI J000503.75+214506.1 & 1.26562 & 21.75171 & 5.36 & 19.76$\pm$0.01\\
  DESI J000619.38+133649.6 & 1.58077 & 13.61380 & 4.86 & 21.19$\pm$0.04\\
  DESI J000918.00--101723.0 & 2.32500 & --10.28972 & 5.33 & 21.27$\pm$0.04\\
  DESI J001040.04+115823.4 & 2.66682 & 11.97319 & 5.44 & 20.17$\pm$0.01\\
  DESI J001149.47+074520.5 & 2.95614 & 7.75570 & 4.92 & 19.61$\pm$0.01\\
  DESI J001744.71+230131.9 & 4.43631 & 23.02553 & 4.77 & 21.38$\pm$0.04\\
  DESI J001835.00+081559.4 & 4.64585 & 8.26652 & 5.34 & 21.05$\pm$0.03\\
  DESI J001912.13+043551.3 & 4.80056 & 4.59760 & 5.03 & 20.20$\pm$0.01\\
  DESI J002633.47+162937.5 & 6.63945 & 16.49376 & 5.10 & 21.35$\pm$0.05\\
  ... & ... & ...& ... & ...  \\
  DESI J231627.96+213737.9 & 349.11649 & 21.62721 & 5.03 & 20.50$\pm$0.02\\
  DESI J231630.51--012428.4 & 349.12711 & --1.40790 & 5.20 & 20.74$\pm$0.01\\
  DESI J232332.06+023848.4 & 350.88361 & 2.64679 & 5.11 & 20.94$\pm$0.03\\
  DESI J232911.44+021720.6 & 352.29767 & 2.28907 & 4.84 & 20.24$\pm$0.01\\
  DESI J233354.68--060522.9 & 353.47784 & --6.08972 & 4.92 & 20.99$\pm$0.03\\
  DESI J233410.03--081702.3 & 353.54178 & --8.28398 & 5.42 & 20.84$\pm$0.03\\
  DESI J233419.15--062626.2 & 353.57980 & --6.44063 & 5.08 & 21.16$\pm$0.04\\
  DESI J233456.62--105351.5 & 353.73593 & --10.89766 & 5.17 & 20.57$\pm$0.02\\
  DESI J233507.31+092725.3 & 353.78046 & 9.45705 & 5.20 & 21.26$\pm$0.03\\
  DESI J234216.88+050856.7 & 355.57034 & 5.14911 & 6.00 & 21.10$\pm$0.02\\
  DESI J235750.51--114244.5 & 359.46045 & --11.71237 & 4.99 & 19.64$\pm$0.01\\
  DESI J235839.43+124708.8 & 359.66428 & 12.78578 & 4.64 & 20.89$\pm$0.02\\
 \enddata
\tablenotetext{a}{Redshift from visual fitting using quasar template with a typical uncer-\\tainty of 0.03. For strong BAL quasars and WL quasars, the uncertainty\\ could be $\sim 0.05-0.1$.}
\tablenotetext{b}{Quasars J012235.47--003602.4 and J102047.40+042946.8 have also been\\ independently discovered in \cite{matsuoka22}.}
\tablenotetext{c}{Quasars J020033.31--173726.0, J180546.94+491824.1, J184520.68+534547.1,\\ and J191435.44+631452.4 have also been  independently discovered in\\ \cite{banados23}.}
\tablenotetext{*}{\textbf{The full table is available in machine-readable format in the \\online version of this article.}}
\label{tab:newquasar}
\end{deluxetable}

\figsetstart
\figsetnum{7}
\figsettitle{The single-exposure spectra of 412 new quasars}

\figsetgrpstart
\figsetgrpnum{7.1}
\figsetgrptitle{Page 1}
\figsetplot{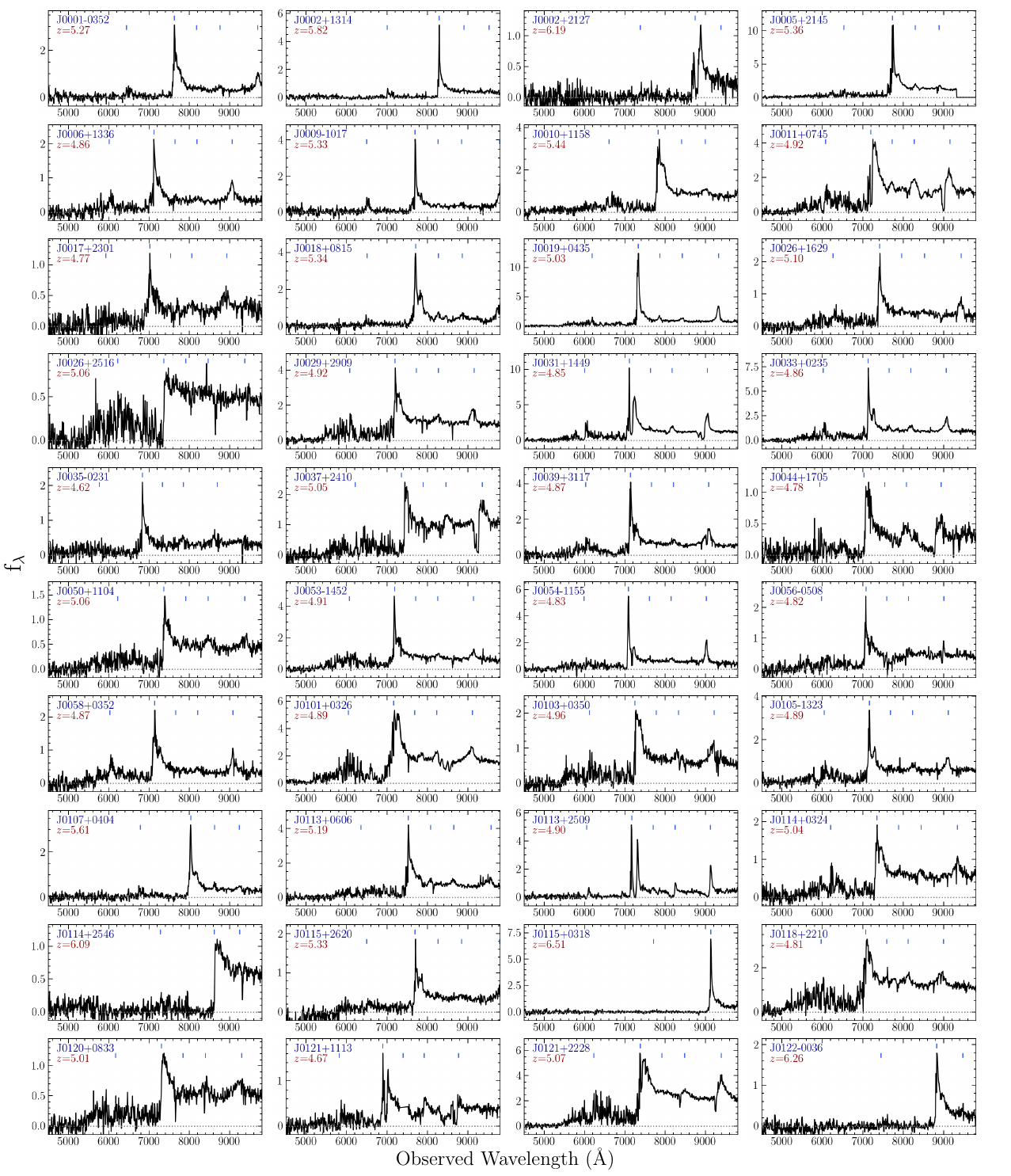}
\figsetgrpnote{The single-exposure spectra of new quasars ordered as in Table \ref{tab:newquasar}, with an actual exposure time of $\sim$ 500s -- 1800s. The flux density ($f_{\rm \lambda}$) is in units of $\rm 10^{-17}~erg~s^{-1}~cm^{-2}~\AA^{-1}$. The spectra have been binned with 11 pixels for the purpose of plotting. The blue vertical lines denote the observed wavelengths of the emission lines, including (from left to right) Ly$\beta$, Ly$\alpha$, O\,{\sc i}, Si\,{\sc iv}, and C\,{\sc iv}.}
\figsetgrpend

\figsetgrpstart
\figsetgrpnum{7.2}
\figsetgrptitle{Page 2}
\figsetplot{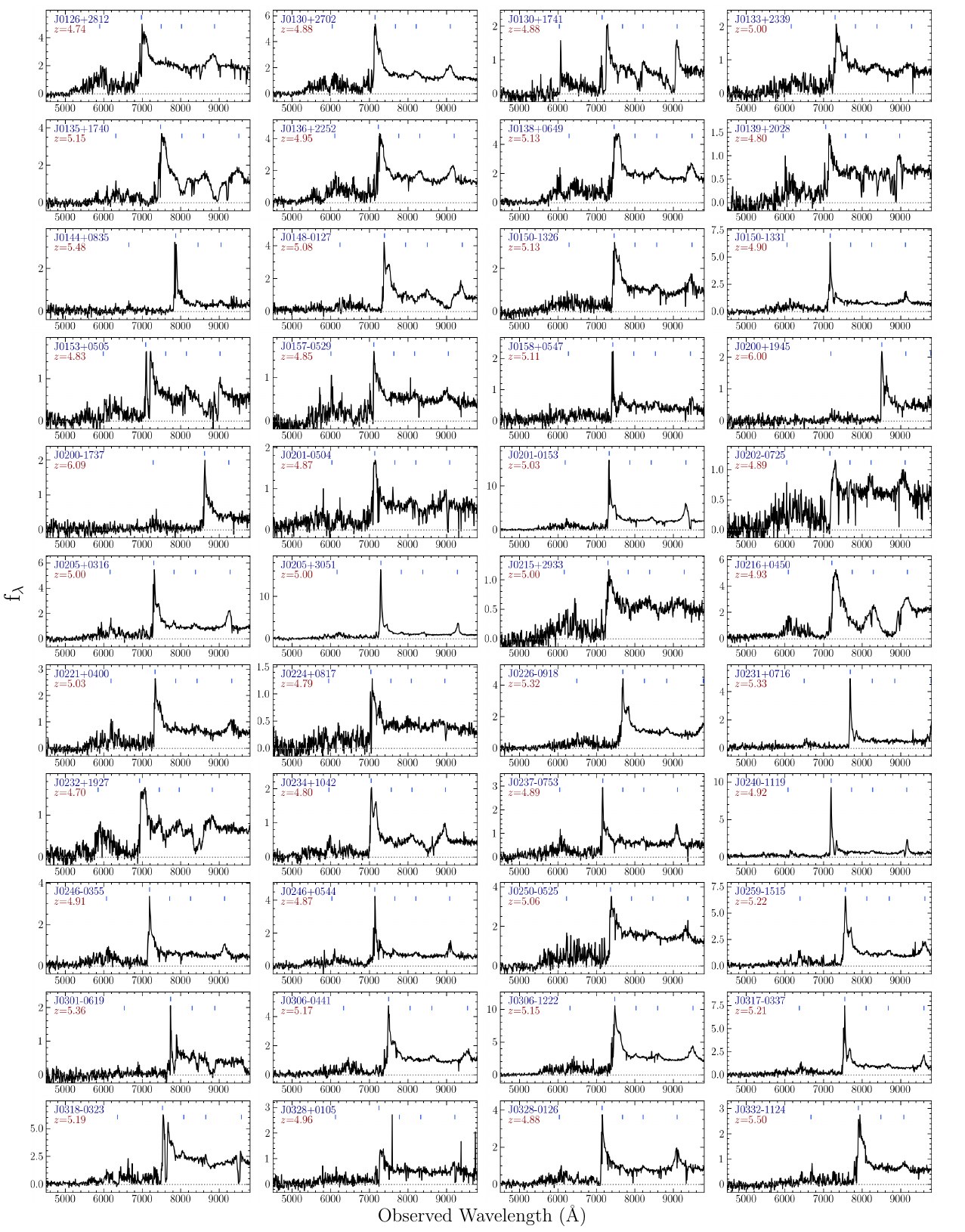}
\figsetgrpnote{The single-exposure spectra of new quasars ordered as in Table \ref{tab:newquasar}, with an actual exposure time of $\sim$ 500s -- 1800s. The flux density ($f_{\rm \lambda}$) is in units of $\rm 10^{-17}~erg~s^{-1}~cm^{-2}~\AA^{-1}$. The spectra have been binned with 11 pixels for the purpose of plotting. The blue vertical lines denote the observed wavelengths of the emission lines, including (from left to right) Ly$\beta$, Ly$\alpha$, O\,{\sc i}, Si\,{\sc iv}, and C\,{\sc iv}.}
\figsetgrpend

\figsetgrpstart
\figsetgrpnum{7.3}
\figsetgrptitle{Page 3}
\figsetplot{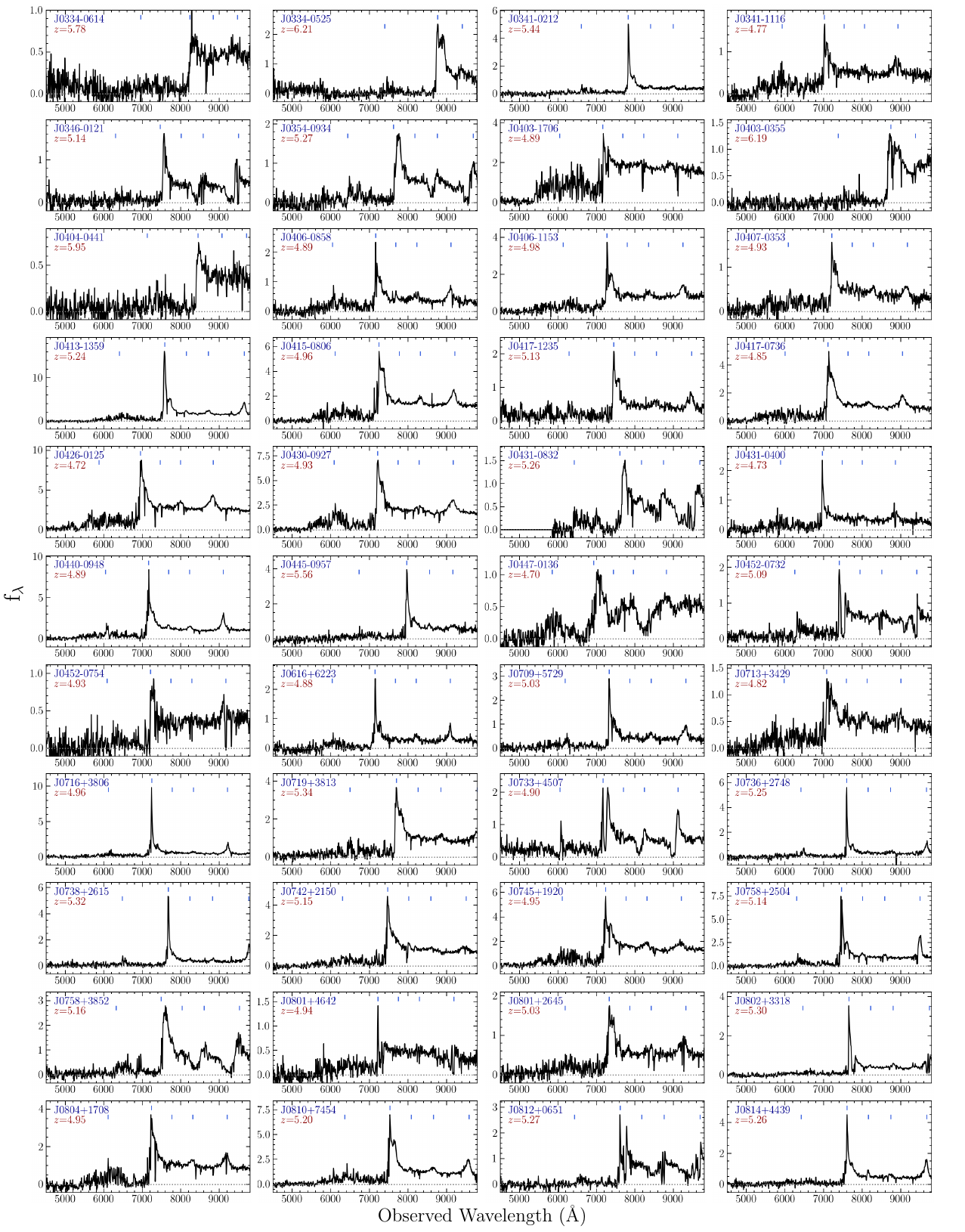}
\figsetgrpnote{The single-exposure spectra of new quasars ordered as in Table \ref{tab:newquasar}, with an actual exposure time of $\sim$ 500s -- 1800s. The flux density ($f_{\rm \lambda}$) is in units of $\rm 10^{-17}~erg~s^{-1}~cm^{-2}~\AA^{-1}$. The spectra have been binned with 11 pixels for the purpose of plotting. The blue vertical lines denote the observed wavelengths of the emission lines, including (from left to right) Ly$\beta$, Ly$\alpha$, O\,{\sc i}, Si\,{\sc iv}, and C\,{\sc iv}.}
\figsetgrpend

\figsetgrpstart
\figsetgrpnum{7.4}
\figsetgrptitle{Page 4}
\figsetplot{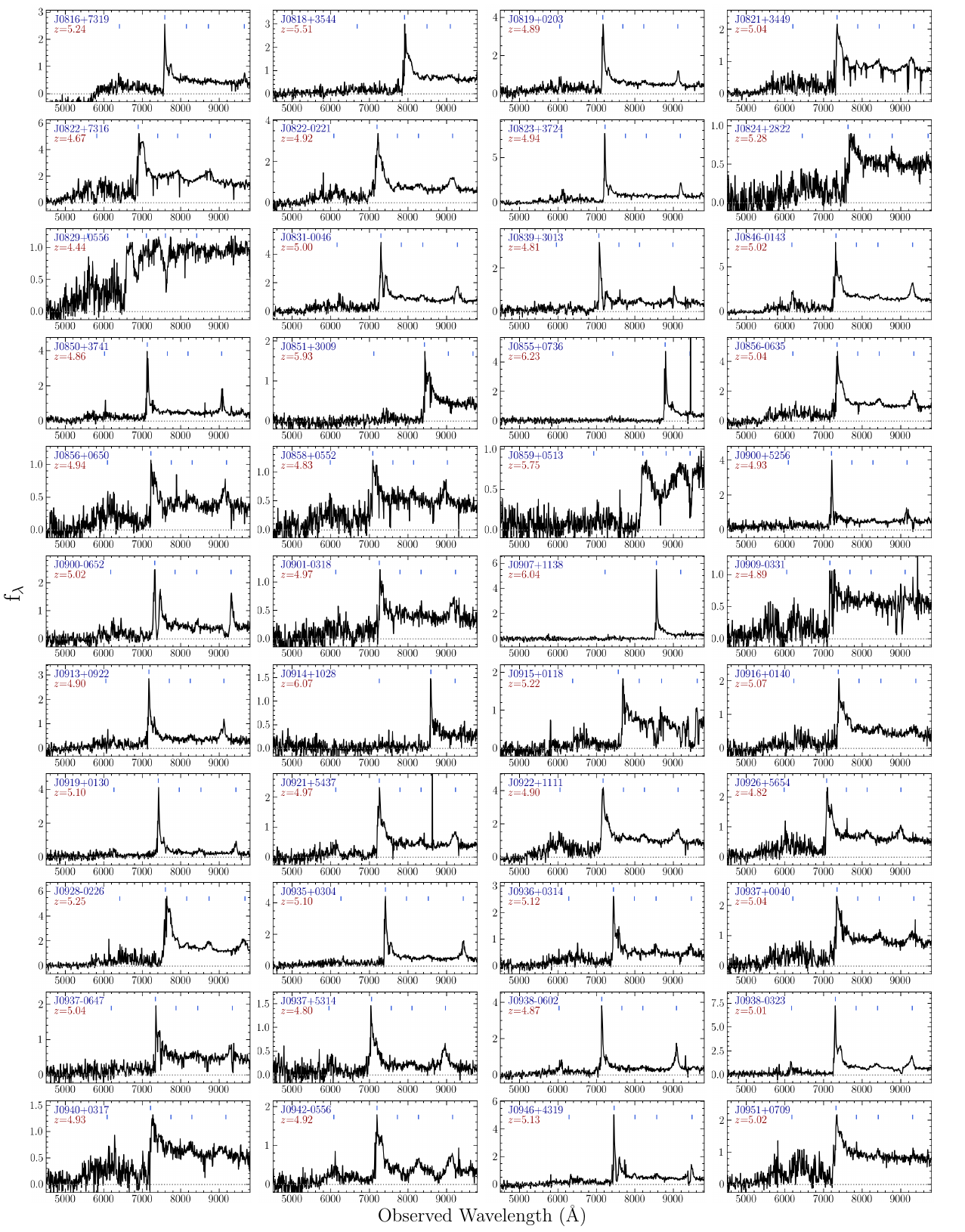}
\figsetgrpnote{The single-exposure spectra of new quasars ordered as in Table \ref{tab:newquasar}, with an actual exposure time of $\sim$ 500s -- 1800s. The flux density ($f_{\rm \lambda}$) is in units of $\rm 10^{-17}~erg~s^{-1}~cm^{-2}~\AA^{-1}$. The spectra have been binned with 11 pixels for the purpose of plotting. The blue vertical lines denote the observed wavelengths of the emission lines, including (from left to right) Ly$\beta$, Ly$\alpha$, O\,{\sc i}, Si\,{\sc iv}, and C\,{\sc iv}.}
\figsetgrpend

\figsetgrpstart
\figsetgrpnum{7.5}
\figsetgrptitle{Page 5}
\figsetplot{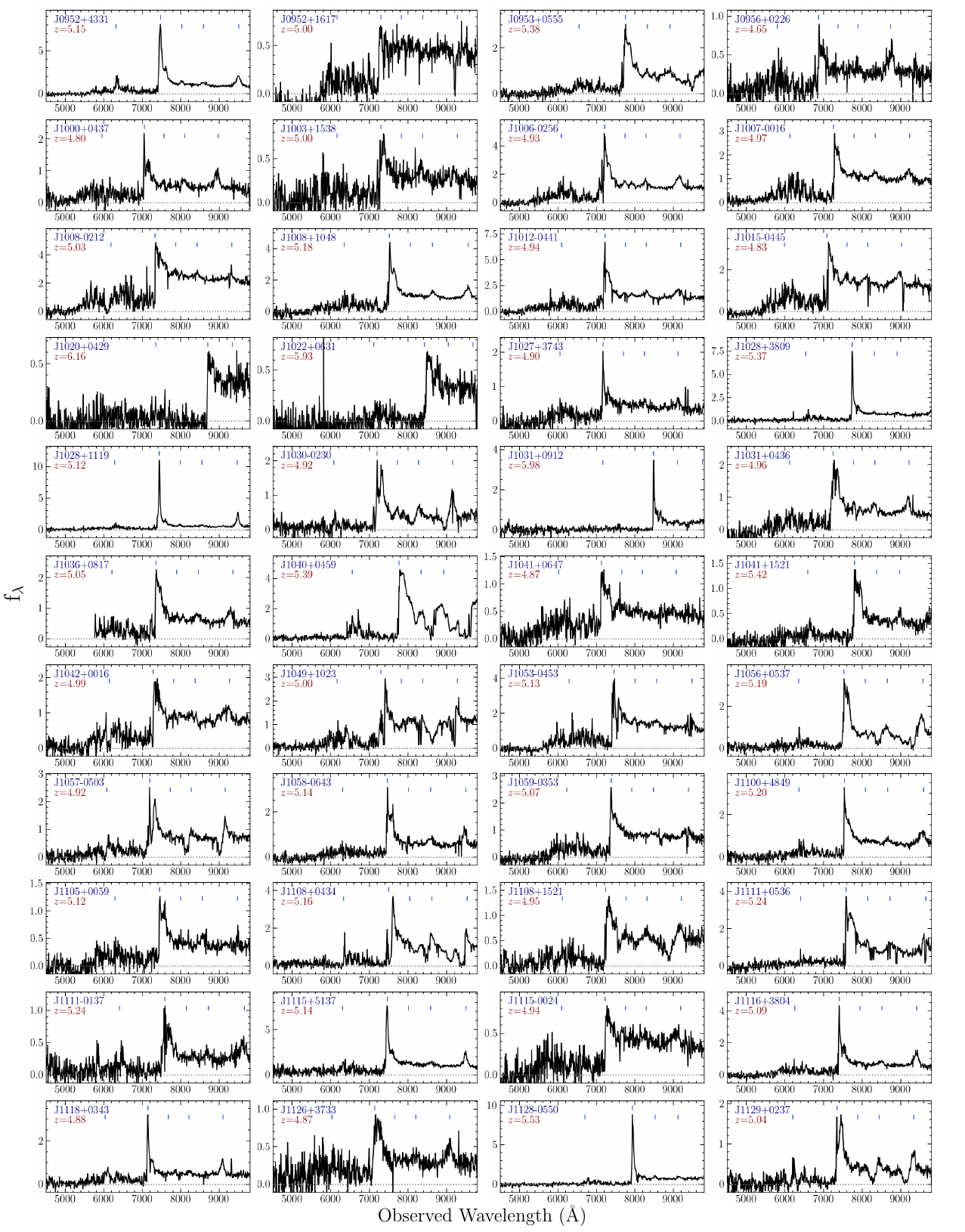}
\figsetgrpnote{The single-exposure spectra of new quasars ordered as in Table \ref{tab:newquasar}, with an actual exposure time of $\sim$ 500s -- 1800s. The flux density ($f_{\rm \lambda}$) is in units of $\rm 10^{-17}~erg~s^{-1}~cm^{-2}~\AA^{-1}$. The spectra have been binned with 11 pixels for the purpose of plotting. The blue vertical lines denote the observed wavelengths of the emission lines, including (from left to right) Ly$\beta$, Ly$\alpha$, O\,{\sc i}, Si\,{\sc iv}, and C\,{\sc iv}.}
\figsetgrpend

\figsetgrpstart
\figsetgrpnum{7.6}
\figsetgrptitle{Page 6}
\figsetplot{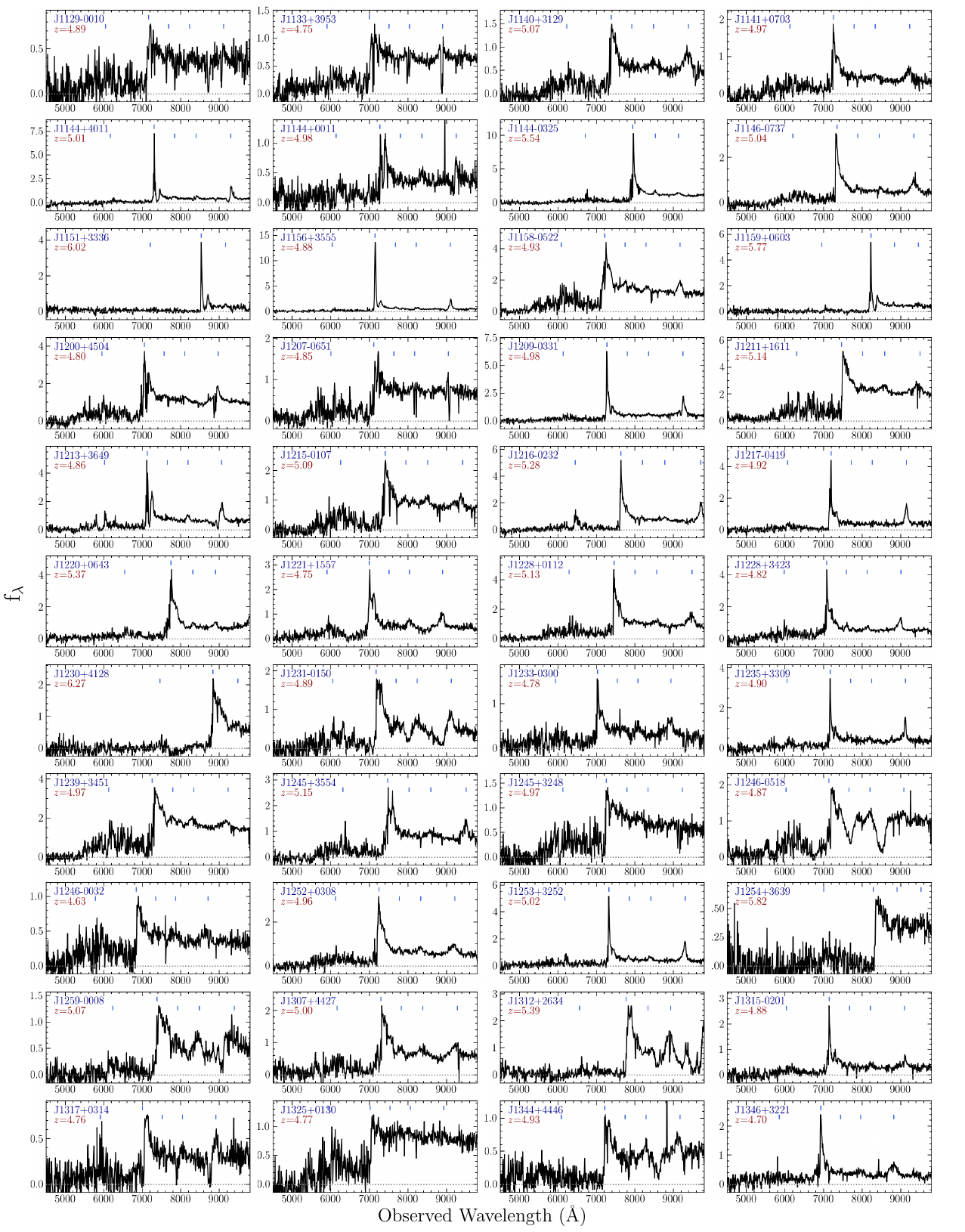}
\figsetgrpnote{The single-exposure spectra of new quasars ordered as in Table \ref{tab:newquasar}, with an actual exposure time of $\sim$ 500s -- 1800s. The flux density ($f_{\rm \lambda}$) is in units of $\rm 10^{-17}~erg~s^{-1}~cm^{-2}~\AA^{-1}$. The spectra have been binned with 11 pixels for the purpose of plotting. The blue vertical lines denote the observed wavelengths of the emission lines, including (from left to right) Ly$\beta$, Ly$\alpha$, O\,{\sc i}, Si\,{\sc iv}, and C\,{\sc iv}.}
\figsetgrpend

\figsetgrpstart
\figsetgrpnum{7.7}
\figsetgrptitle{Page 7}
\figsetplot{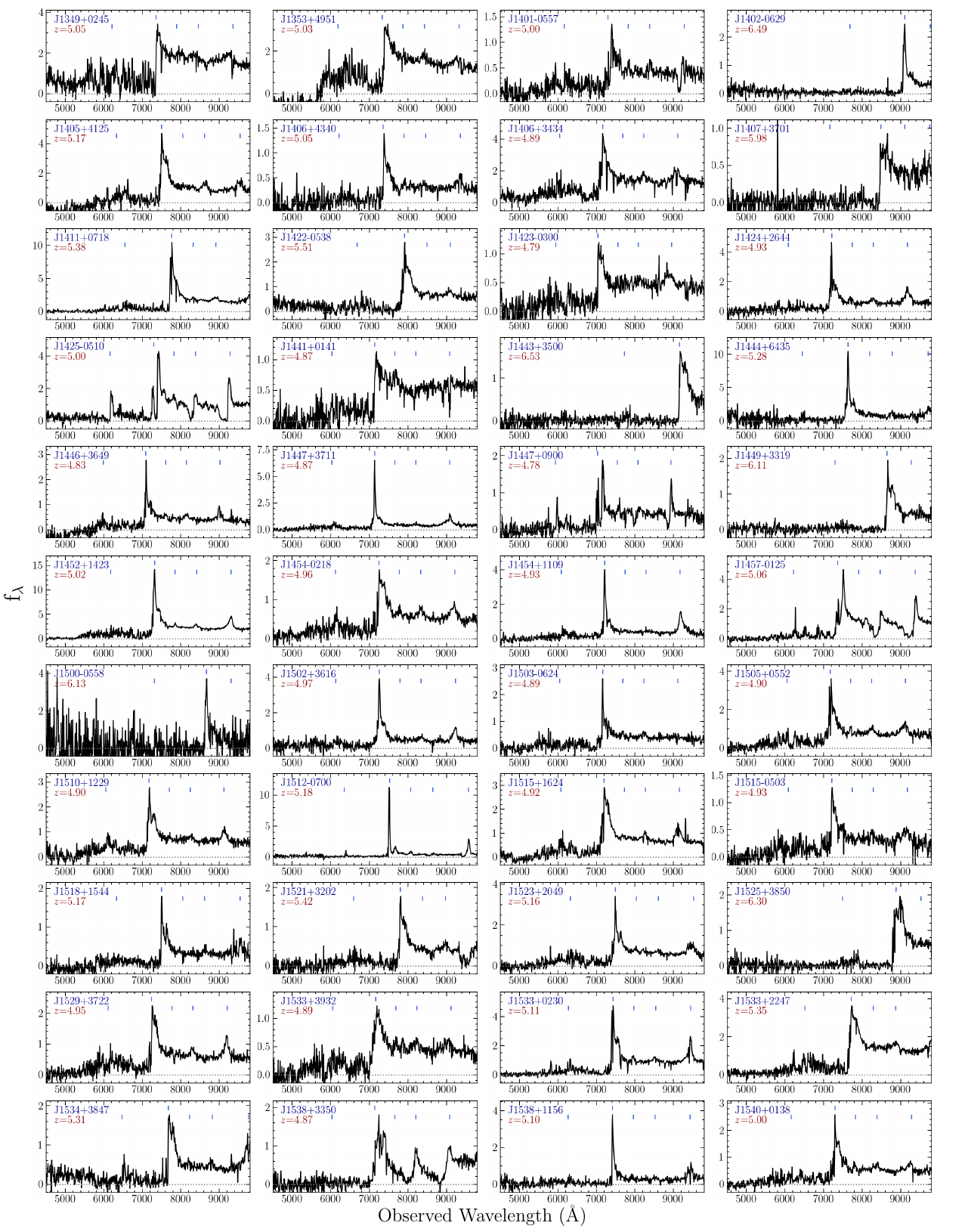}
\figsetgrpnote{The single-exposure spectra of new quasars ordered as in Table \ref{tab:newquasar}, with an actual exposure time of $\sim$ 500s -- 1800s. The flux density ($f_{\rm \lambda}$) is in units of $\rm 10^{-17}~erg~s^{-1}~cm^{-2}~\AA^{-1}$. The spectra have been binned with 11 pixels for the purpose of plotting. The blue vertical lines denote the observed wavelengths of the emission lines, including (from left to right) Ly$\beta$, Ly$\alpha$, O\,{\sc i}, Si\,{\sc iv}, and C\,{\sc iv}.}
\figsetgrpend

\figsetgrpstart
\figsetgrpnum{7.8}
\figsetgrptitle{Page 8}
\figsetplot{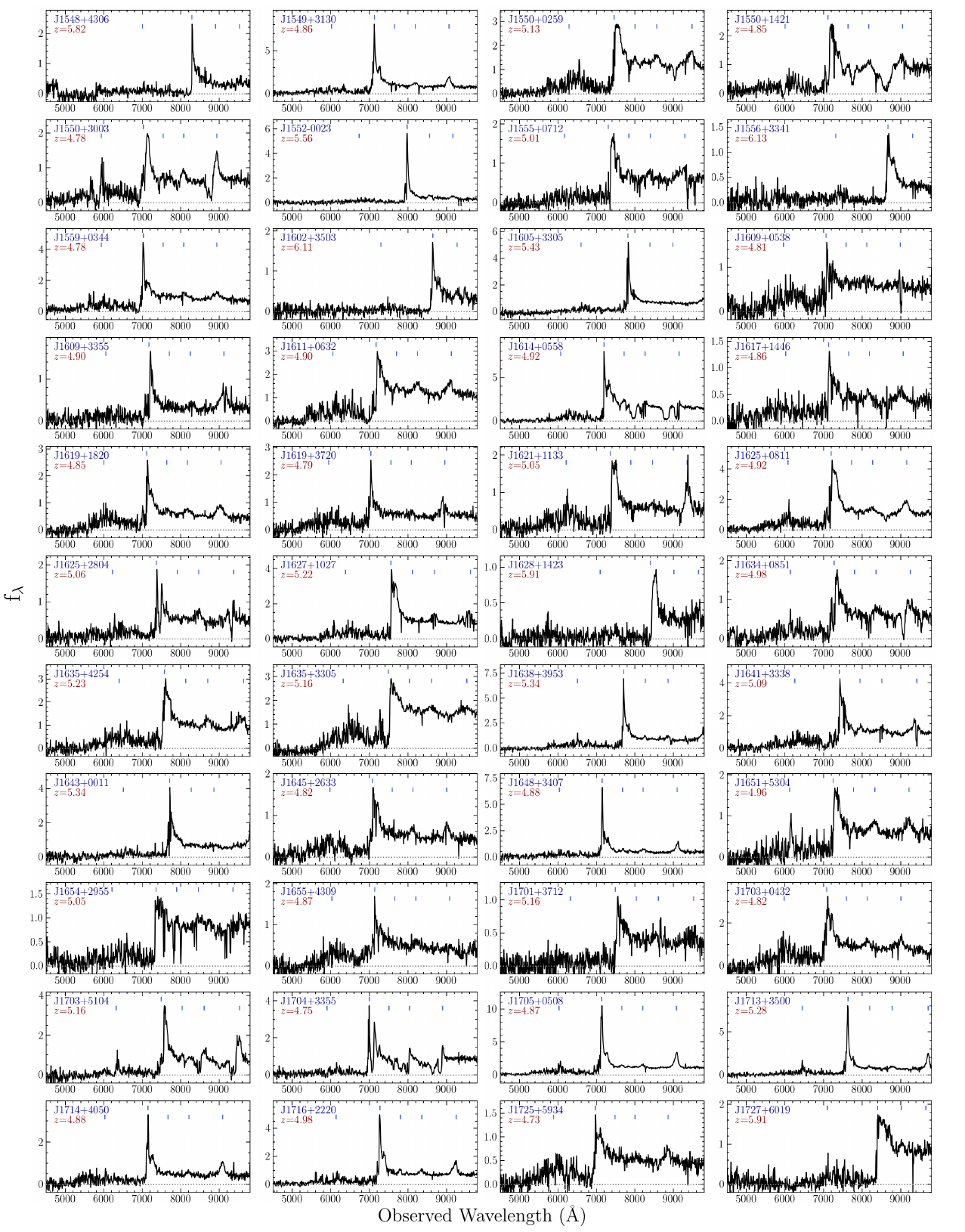}
\figsetgrpnote{The single-exposure spectra of new quasars ordered as in Table \ref{tab:newquasar}, with an actual exposure time of $\sim$ 500s -- 1800s. The flux density ($f_{\rm \lambda}$) is in units of $\rm 10^{-17}~erg~s^{-1}~cm^{-2}~\AA^{-1}$. The spectra have been binned with 11 pixels for the purpose of plotting. The blue vertical lines denote the observed wavelengths of the emission lines, including (from left to right) Ly$\beta$, Ly$\alpha$, O\,{\sc i}, Si\,{\sc iv}, and C\,{\sc iv}.}
\figsetgrpend

\figsetgrpstart
\figsetgrpnum{7.9}
\figsetgrptitle{Page 9}
\figsetplot{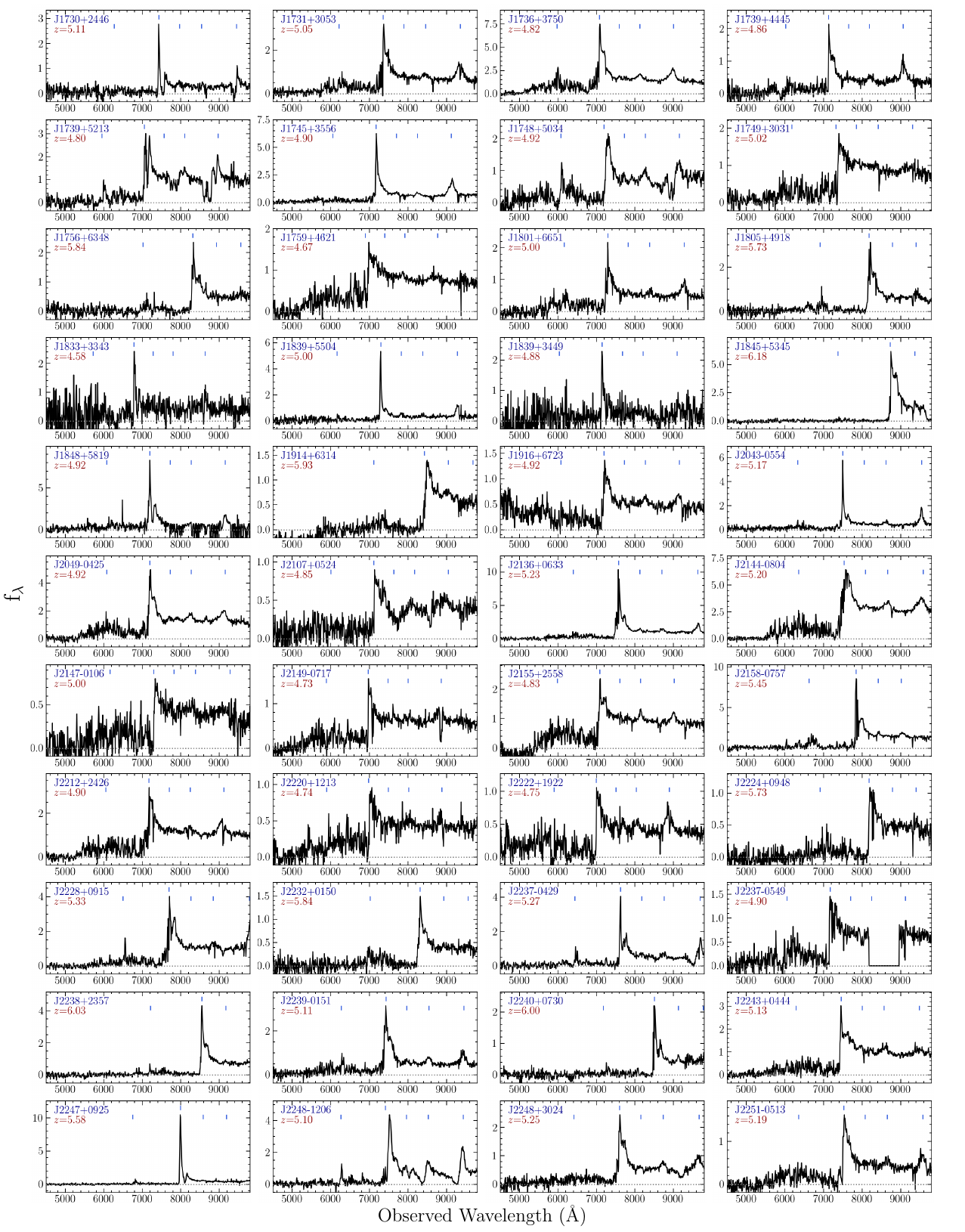}
\figsetgrpnote{The single-exposure spectra of new quasars ordered as in Table \ref{tab:newquasar}, with an actual exposure time of $\sim$ 500s -- 1800s. The flux density ($f_{\rm \lambda}$) is in units of $\rm 10^{-17}~erg~s^{-1}~cm^{-2}~\AA^{-1}$. The spectra have been binned with 11 pixels for the purpose of plotting. The blue vertical lines denote the observed wavelengths of the emission lines, including (from left to right) Ly$\beta$, Ly$\alpha$, O\,{\sc i}, Si\,{\sc iv}, and C\,{\sc iv}.}
\figsetgrpend

\figsetgrpstart
\figsetgrpnum{7.10}
\figsetgrptitle{Page 10}
\figsetplot{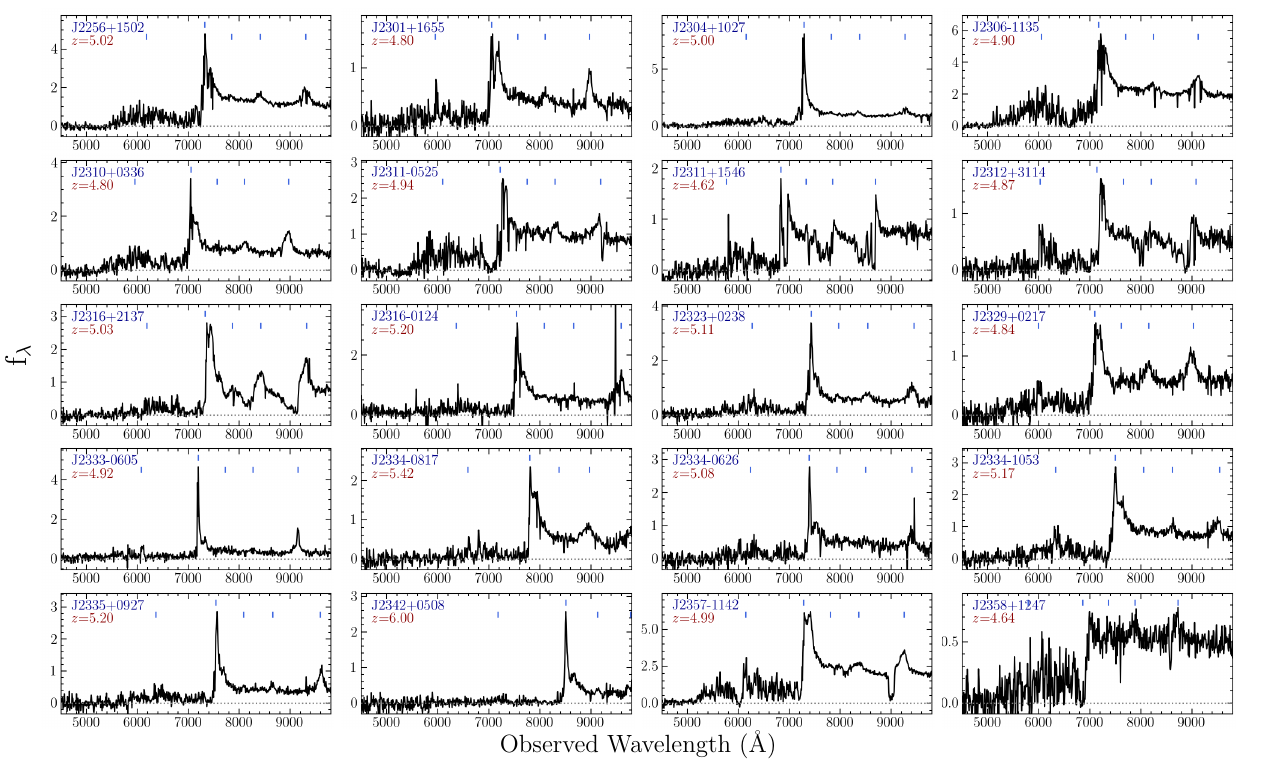}
\figsetgrpnote{The single-exposure spectra of new quasars ordered as in Table \ref{tab:newquasar}, with an actual exposure time of $\sim$ 500s -- 1800s. The flux density ($f_{\rm \lambda}$) is in units of $\rm 10^{-17}~erg~s^{-1}~cm^{-2}~\AA^{-1}$. The spectra have been binned with 11 pixels for the purpose of plotting. The blue vertical lines denote the observed wavelengths of the emission lines, including (from left to right) Ly$\beta$, Ly$\alpha$, O\,{\sc i}, Si\,{\sc iv}, and C\,{\sc iv}.}
\figsetgrpend

\figsetend

\begin{figure*}%
\centering 
\epsscale{1.22}
\plotone{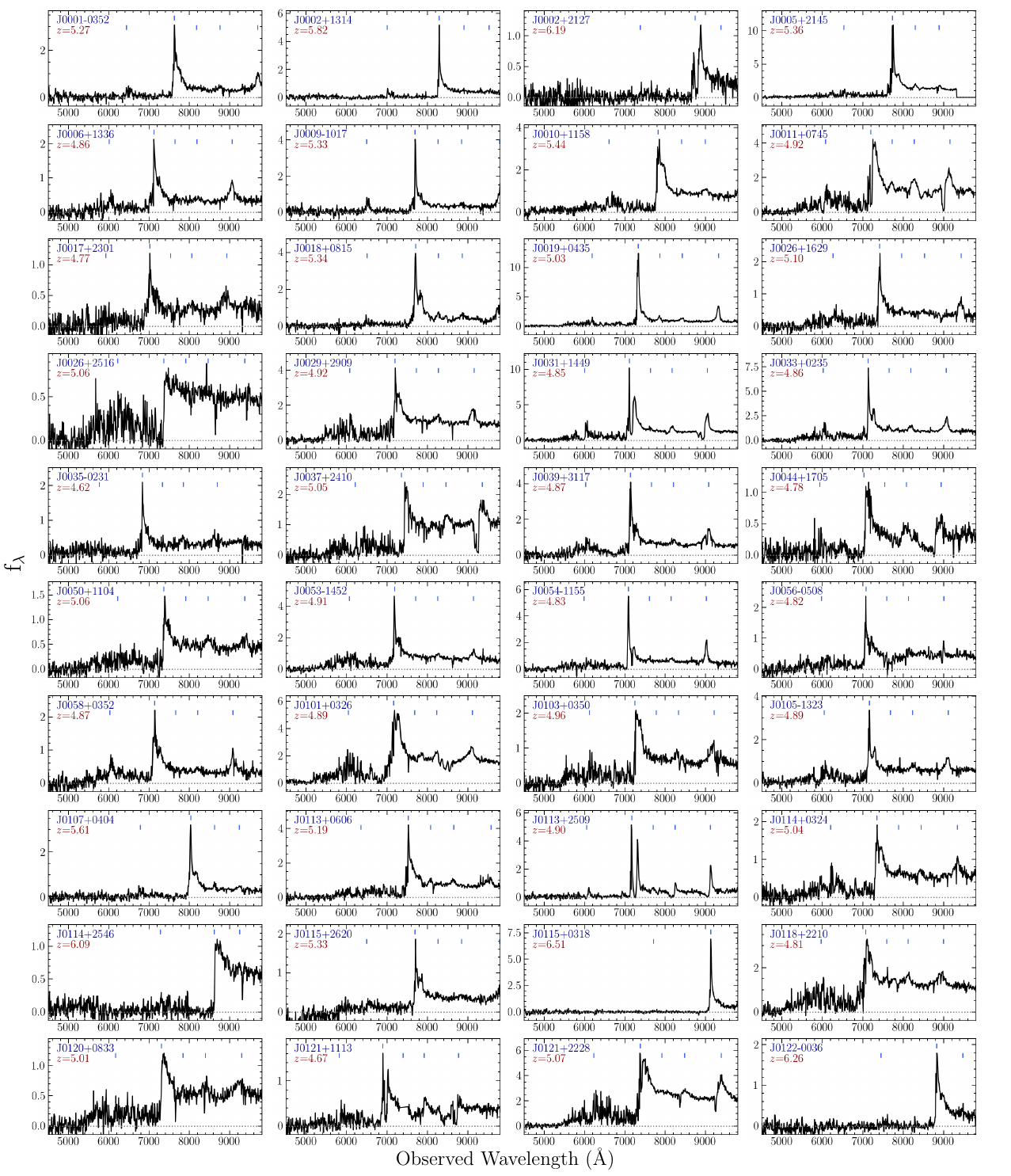} 
\vspace{-25pt}
\caption{The single-exposure spectra of new quasars ordered as in Table \ref{tab:newquasar}, with an actual exposure time of $\sim$ 500s -- 1800s. The flux density ($f_{\rm \lambda}$) is in units of $\rm 10^{-17}~erg~s^{-1}~cm^{-2}~\AA^{-1}$. The spectra have been binned with 11 pixels for the purpose of plotting. The blue vertical lines denote the observed wavelengths of the emission lines, including (from left to right) Ly$\beta$, Ly$\alpha$, O\,{\sc i}, Si\,{\sc iv}, and C\,{\sc iv}. We are presenting the first two pages of the spectral sample, and the complete figure set (10 pages, 412 spectra) is available in the online journal.}
\label{fig:allspectra}
\end{figure*}
 
\addtocounter{figure}{-1}
\begin{figure*}%
\centering 
\epsscale{1.22}
\plotone{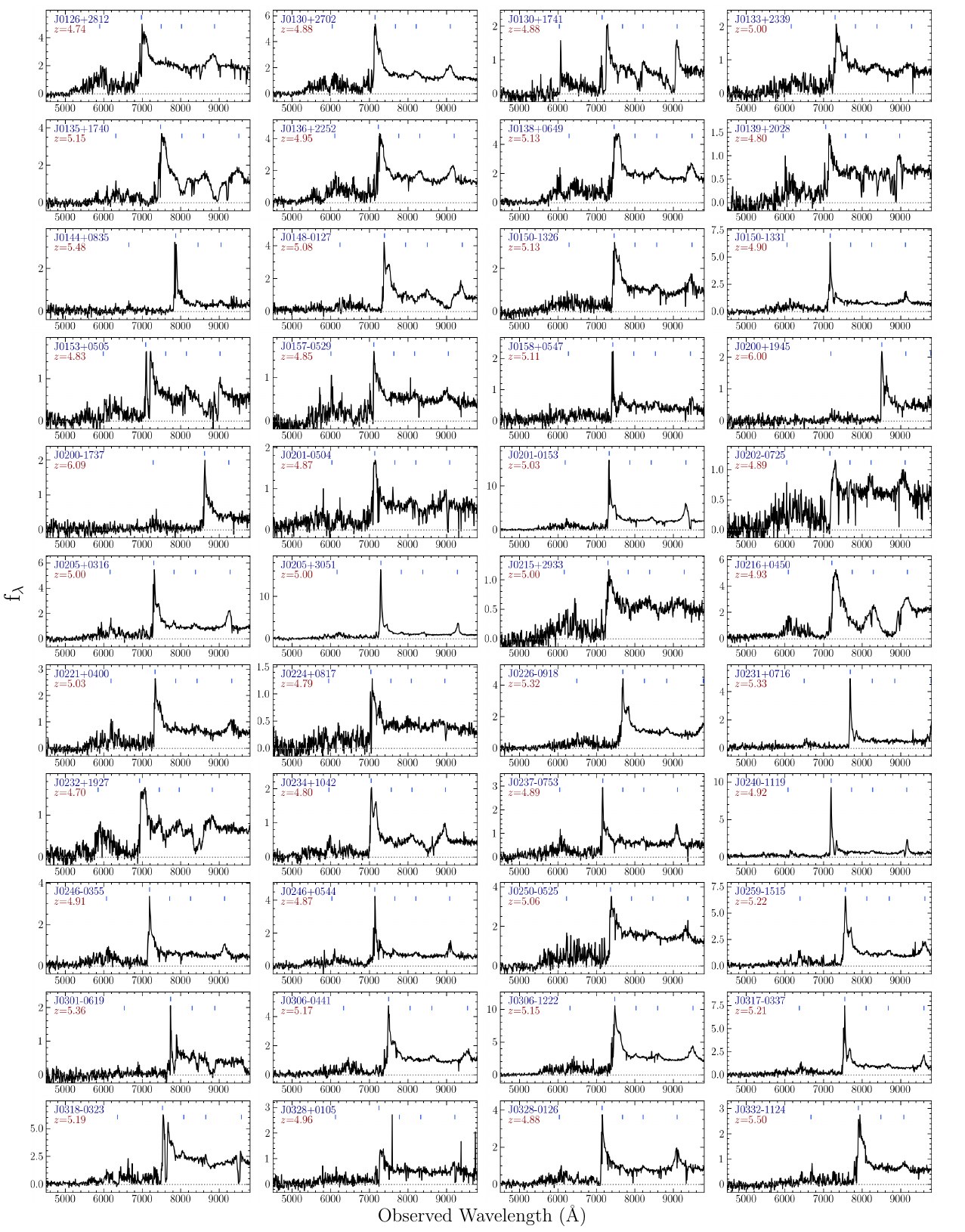} 
\caption{Continued. We are presenting the first two pages of the spectral sample, and the complete figure set (10 pages, 412 spectra) is available in the online journal.}
\label{fig:allspectra02}
\end{figure*}

\clearpage
\acknowledgments
We thank Alex Kim, Robert Knop, and Divij Sharma for the DESI internal database.
X. Fan acknowledges support from US NSF grants AST 15-15115, AST 19-08284 and NASA ADAP Grant NNX17AF28G.
ADM was supported by the U.S. Department of Energy, Office of Science, Office of High Energy Physics, under Award Number DE-SC0019022.
F. Wang acknowledges support by NASA through the NASA Hubble Fellowship grant $\#$HST-HF2-51448.001-A and $\#$HF2-51434 awarded by the Space Telescope Science Institute, which is operated by the Association of Universities for Research in Astronomy, Incorporated, under NASA contract NAS5-26555. 

This material is based upon work supported by the U.S. Department of Energy (DOE), Office of Science, Office of High-Energy Physics, under Contract No. DE-AC02-05CH11231, and by the National Energy Research Scientific Computing Center, a DOE Office of Science User Facility under the same contract. Additional support for DESI was provided by the U.S. National Science Foundation (NSF), Division of Astronomical Sciences under Contract No. AST-0950945 to the NSF's National Optical-Infrared Astronomy Research Laboratory; the Science and Technology Facilities Council of the United Kingdom; the Gordon and Betty Moore Foundation; the Heising-Simons Foundation; the French Alternative Energies and Atomic Energy Commission (CEA); the National Council of Science and Technology of Mexico (CONACYT); the Ministry of Science and Innovation of Spain (MICINN), and by the DESI Member Institutions: https://www.desi.lbl.gov/collaborating-institutions.

The DESI Legacy Imaging Surveys consist of three individual and complementary projects: the Dark Energy Camera Legacy Survey (DECaLS), the Beijing-Arizona Sky Survey (BASS), and the Mayall z-band Legacy Survey (MzLS). DECaLS, BASS and MzLS together include data obtained, respectively, at the Blanco telescope, Cerro Tololo Inter-American Observatory, NSF?s NOIRLab; the Bok telescope, Steward Observatory, University of Arizona; and the Mayall telescope, Kitt Peak National Observatory, NOIRLab. NOIRLab is operated by the Association of Universities for Research in Astronomy (AURA) under a cooperative agreement with the National Science Foundation. Pipeline processing and analyses of the data were supported by NOIRLab and the Lawrence Berkeley National Laboratory. Legacy Surveys also uses data products from the Near-Earth Object Wide-field Infrared Survey Explorer (NEOWISE), a project of the Jet Propulsion Laboratory/California Institute of Technology, funded by the National Aeronautics and Space Administration. Legacy Surveys was supported by: the Director, Office of Science, Office of High Energy Physics of the U.S. Department of Energy; the National Energy Research Scientific Computing Center, a DOE Office of Science User Facility; the U.S. National Science Foundation, Division of Astronomical Sciences; the National Astronomical Observatories of China, the Chinese Academy of Sciences and the Chinese National Natural Science Foundation. LBNL is managed by the Regents of the University of California under contract to the U.S. Department of Energy. The complete acknowledgments can be found at https://www.legacysurvey.org/.

Any opinions, findings, and conclusions or recommendations expressed in this material are those of the author(s) and do not necessarily reflect the views of the U. S. National Science Foundation, the U. S. Department of Energy, or any of the listed funding agencies.

The authors are honored to be permitted to conduct scientific research on Iolkam Du'ag (Kitt Peak), a mountain with particular significance to the Tohono O'odham Nation.

%

\vspace{5mm}
\facilities{Mayall (DESI)}, {Blanco (DECam)}, {Mayall (Mosaic-3)}, {Bok (90Prime)}, {WISE}, {Pan-STARR1 (GPC1)}, {UKIRT (WFCam)}, {VISTA (VIRCAM)}





\appendix
\section{Selection Criteria for Survey of quasars at $z \sim 4.8-6.8$}\label{A}
The Survey of quasars at $z \sim 4.8-6.8$ has been described in Section 2.2. Here we present the full color-color selection criteria for each redshift range. The $J,W1$ and $W2$ magnitudes are used in the Vega systems ($J_{\rm AB}$ = $J_{\rm Vega} + 0.938$, $W1_{\rm AB} = W1_{\rm Vega} + 2.699$, $W2_{\rm AB} = W2_{\rm Vega} + 3.339$).  We performed a pre-selection to reduce the candidate sample size using the following criteria: 

\begin{equation}
\begin{array}{l}
{\rm S/N}(z) > 5 ~{\rm and}~ z < 21.5;\\
{\rm S/N}(W1) > 3 ~{\rm and}~ {\rm S/N}(W2) > 2;\\
{\rm S/N}(g) < 5~{\rm or}~g > 24.5 ~{\rm or}~g-r > 1.8;\\
{\rm S/N}(r) < 5 ~{\rm or}~ r - z >1.0
\end{array}
\end{equation}

We then matched the pre-selected sample with the PS1 photometric catalogs and applied color selections to select quasar candidates for different redshift ranges.\\

\noindent$z \sim 4.8-5.4$ quasar candidates:
\begin{equation}
\begin{array}{l}
{\rm S/N}(g) < 3 ~{\rm or}~ g - r > 2.5;\\
{\rm S/N}(i_{\rm P1}) > 5 ~{\rm and}~ z < 21.4;\\
r-i_{\rm P1}>0.8 ~{\rm and}~ r-i_{\rm P1}<3.0;\\
i_{\rm P1}-z<0.5\times(r-i_{\rm P1})-0.2;\\
i_{\rm P1}-z>-0.5 ~{\rm and}~ i_{\rm P1}-z<0.7;\\
z - W1 > 2.5 ~{\rm and}~ W1 - W2 >0.5;\\
\text{If objects have PS1 $y_{\rm P1}$ photometry,}\\
{\rm S/N}(y_{\rm P1}) < 3 ~{\rm or}~ ( y_{\rm P1}-W1> 2.399 ~{\rm and}~ z-y_{\rm P1}<0.5)
\end{array}
\end{equation}

\noindent$z \sim 5.7-6.4$ quasar candidates:
\begin{equation}
\begin{array}{l}
{\rm S/N}(g) < 3 ~{\rm or}~ g > 24.8;\\
{\rm S/N}(r) < 3 ~{\rm or}~ r > 24 ~{\rm or}~ r - z > 3;\\
{\rm S/N}(i_{\rm P1}) < 3 ~{\rm or}~ i_{\rm P1} - z > 2.0; \\
z - W1 > 2.5 ~{\rm and}~ W1 - W2 >0.5;\\
\text{If objects have PS1 photometry,}\\
{\rm S/N}(z_{\rm P1})<3 ~{\rm or}~  {\rm S/N}(y_{\rm P1}) < 3 ~{\rm or}~ \\
(({\rm S/N}(i_{\rm P1})<3 ~{\rm or}~ i_{\rm P1}-z_{\rm P1}>2.0) ~{\rm and}~ z_{\rm P1}-y_{\rm P1}<1.0);\\
{\rm S/N}(y_{\rm P1}) < 3 ~{\rm or}~ y_{\rm P1}-W1> 2.399
\end{array}
\end{equation}

\noindent$z > 6.4$ quasar candidates:
\begin{equation}
\begin{array}{l}
{\rm S/N}(g) < 3 ~{\rm or}~ g>25.0;\\
{\rm S/N}(r) < 3 ~{\rm or}~ r>25.0;\\
{\rm S/N}(i_{\rm P1}) < 3 ~{\rm or}~ i_{\rm P1}>23.5 ~{\rm or}~ i_{\rm P1} - y_{\rm P1}>3.0;\\
z-W1 > 2.799 ~{\rm and}~ W1- W2> 0.5;\\
{\rm S/N}(y_{\rm P1}) > 3;\\
y_{\rm P1} -W1>2.399 ~{\rm and}~ z-y_{\rm P1}>0.4;\\
\text{If objects have $z_{\rm P1}$ photometry,}\\
{\rm S/N}(z_{\rm P1})< 3 ~{\rm or}~ (z_{\rm P1}-y_{\rm P1}>1.0 ~{\rm and}~ z_{\rm P1}-W1>4.0)
\end{array}
\end{equation}

Then, for objects with $J-$band detection ($\rm S/N(J) > 3$), we applied $J-$band related colors to further reject contaminants by removing objects with the following colors:
\begin{equation}
\begin{array}{l}
y_{\rm P1} - J>2.0 ~{\rm or}~ J-W1<1.5
\end{array}
\end{equation}

\section{Selection of $z\sim 4- 5.3$ Quasars in SV1} \label{B}
\begin{figure*}%
\centering 
\epsscale{1.2}
\plotone{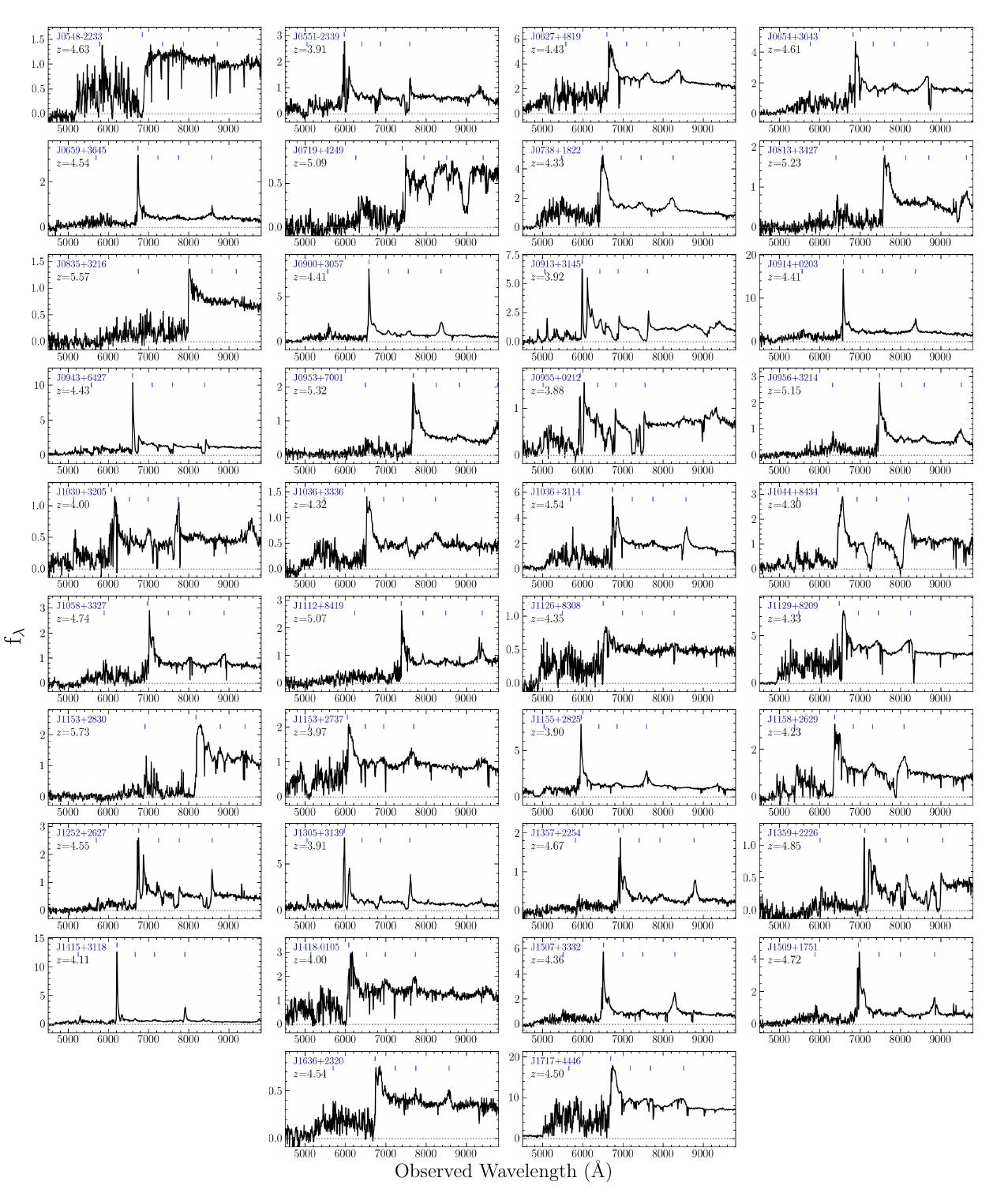} 
\vspace{-15pt}
\caption{The spectra of new quasars from the SV1 selection ordered as in Table \ref{tab:sv1quasar}. The y-axis shows the flux density ($f_{\rm \lambda}$) in units of $\rm 10^{-17}~erg~s^{-1}~cm^{-2}~\AA^{-1}$. The spectra have been binned with 11 pixels for the purpose of plotting. The blue vertical lines denote the observed wavelengths of the emission lines, including (from left to right) Ly$\beta$, Ly$\alpha$, O\,{\sc i}, Si\,{\sc iv}, and C\,{\sc iv}. Quasar J0548--2233 is an example of WL quasar, and J0719+4249 is an example of BAL quasar.}
\label{fig:sv1spectra}
\end{figure*}

In Table \ref{tab:sv1quasar}, we list all 38 new quasars from the selection of $z \sim 4-5.3$ quasars during the DESI SV1 observations. The spectra are shown in Figure \ref{fig:sv1spectra}.
We also list the selection criteria of this pilot SV selection below. 

We first require $g$ drop-out and limit S/N in the $z$, $W1$, and $W2$ bands.
\begin{equation}
\begin{array}{l}
{\rm S/N}(g) < 3~{\rm or}~g > 24.5~{\rm or}~g-r > 1.8; \\
{\rm S/N}(z) > 5; \\
{\rm S/N}(W1) > 3, ~{\rm S/N}(W2) > 2
\end{array}
\end{equation}

Then, we build up different selection criteria for $z\sim 4 - 4.8$ and $z\sim 4.8 - 5.3$ quasars. $W1$ and $W2$ are in the Vega magnitude system. 
\begin{equation}
\begin{array}{l}
z\sim 4.8 - 5.3: \\
W1-W2 >0.5,~z-W1> 2.0,~z - W1 < 4.5; \\
({\rm S/N}(r) <3 ~{\rm or }~ \\
(r-z > 1.0 ~{\rm and }~ r-z <3.9  \\ ~{\rm and }~ r-z < (z-w1)\times3.2 - 6.5) \\
~{\rm or }~ r-z > 4.4); \\
z < 21.4.
\end{array}
\end{equation}

\begin{equation}
\begin{array}{l}
z\sim 4 -4.8: \\
W1-W2 >0.3,~z-W1> 2.5,~z - W1 < 4.5; \\
({\rm S/N}(r) > 3,~r-z > -1.0,~r-z <1.5; \\
z < 21.4.
\end{array}
\end{equation}

\begin{deluxetable}{l c l c c}
\tablecaption{The 38 New Quasars from the Selection in SV1.}
\tabletypesize{\footnotesize}
\tablewidth{\textwidth}
\setlength{\tabcolsep}{3pt}
\tablehead{
\colhead{Name} &
\colhead{RA} &
\colhead{Dec} &
\colhead{$z$\tablenotemark{a}} &
\colhead{$z_{\rm LS}$} 
}
\startdata
  DESI J054840.05--223313.8 & 87.16687 & --22.55383 & 4.63 & 20.38$\pm$0.01\\
  DESI J055103.87--233959.8 & 87.76612 & --23.66664 & 3.91 & 20.72$\pm$0.01\\
  DESI J062712.07+481919.1 & 96.80030 & 48.32199 & 4.43 & 19.25$\pm$0.01\\
  DESI J065435.94+364333.9 & 103.64975 & 36.72609 & 4.61 & 19.66$\pm$0.02\\
  DESI J065918.89+364504.0 & 104.82872 & 36.75113 & 4.54 & 21.15$\pm$0.03\\
  DESI J071943.37+424955.2 & 109.93072 & 42.83202 & 5.09 & 21.21$\pm$0.02\\
  DESI J073853.30+182213.2 & 114.72207 & 18.37034 & 4.33 & 20.11$\pm$0.01\\
  DESI J081344.95+342726.8 & 123.43731 & 34.45746 & 5.23 & 20.68$\pm$0.02\\
  DESI J083558.88+321631.1 & 128.99533 & 32.27533 & 5.57 & 20.65$\pm$0.02\\
  DESI J090025.72+305706.6 & 135.10716 & 30.95184 & 4.41 & 20.68$\pm$0.03\\
  DESI J091332.51+314501.0 & 138.38547 & 31.75029 & 3.92 & 20.12$\pm$0.01\\
  DESI J091444.89+020337.8 & 138.68703 & 2.06051 & 4.41 & 19.87$\pm$0.01\\
  DESI J094314.99+642700.9 & 145.81244 & 64.45025 & 4.43 & 20.02$\pm$0.01\\
  DESI J095330.88+700140.8 & 148.37868 & 70.02802 & 5.32 & 21.05$\pm$0.03\\
  DESI J095519.52+021229.7 & 148.83131 & 2.20825 & 3.88 & 20.70$\pm$0.02\\
  DESI J095629.67+321432.7 & 149.12361 & 32.24243 & 5.15 & 21.14$\pm$0.05\\
  DESI J103049.13+320538.7 & 157.70470 & 32.09411 & 4.00 & 21.08$\pm$0.03\\
  DESI J103609.43+333654.0 & 159.03931 & 33.61501 & 4.32 & 21.29$\pm$0.03\\
  DESI J103630.59+311402.5 & 159.12747 & 31.23405 & 4.54 & 19.84$\pm$0.01\\
  DESI J104418.06+843435.9 & 161.07526 & 84.57666 & 4.30 & 19.94$\pm$0.01\\
  DESI J105805.63+332702.6 & 164.52344 & 33.45073 & 4.74 & 20.44$\pm$0.01\\
  DESI J111233.93+841901.5 & 168.14139 & 84.31708 & 5.07 & 20.48$\pm$0.02\\
  DESI J112612.76+830836.4 & 171.55316 & 83.14345 & 4.35 & 21.01$\pm$0.03\\
  DESI J112902.93+820927.5 & 172.26221 & 82.15764 & 4.33 & 18.82$\pm$0.01\\
  DESI J115329.59+283027.1 & 178.37328 & 28.50753 & 5.73 & 19.98$\pm$0.02\\
  DESI J115356.93+273711.5 & 178.48720 & 27.61987 & 3.97 & 20.52$\pm$0.01\\
  DESI J115537.05+282532.7 & 178.90436 & 28.42577 & 3.90 & 20.56$\pm$0.02\\
  DESI J115855.52+262912.6 & 179.73132 & 26.48685 & 4.23 & 20.50$\pm$0.02\\
  DESI J125214.84+262745.9 & 193.06181 & 26.46275 & 4.55 & 21.06$\pm$0.04\\
  DESI J130530.85+313944.1 & 196.37856 & 31.66225 & 3.91 & 20.84$\pm$0.02\\
  DESI J135716.95+225459.6 & 209.32062 & 22.91656 & 4.67 & 21.36$\pm$0.03\\
  DESI J135942.71+222634.2 & 209.92795 & 22.44285 & 4.85 & 21.20$\pm$0.03\\
  DESI J141550.25+311821.3 & 213.95940 & 31.30592 & 4.11 & 21.10$\pm$0.03\\
  DESI J141809.66--010502.3 & 214.54027 & --1.08399 & 4.00 & 19.90$\pm$0.01\\
  DESI J150753.26+333228.9 & 226.97192 & 33.54136 & 4.36 & 20.69$\pm$0.02\\
  DESI J150931.23+175158.6 & 227.38013 & 17.86628 & 4.72 & 20.70$\pm$0.03\\
  DESI J163602.51+232043.0 & 249.01047 & 23.34530 & 4.54 & 21.25$\pm$0.03\\
  DESI J171717.79+444653.3 & 259.32413 & 44.78148 & 4.50 & 18.13$\pm$0.002
 \enddata
 \tablenotetext{a}{Redshift from visual fitting using quasar template with a typical uncertainty of 0.03. For strong BAL quasars and WL quasars, the uncertainty could be $\sim 0.05-0.1$.}
 \tablenotetext{*}{\textbf{A machine-readable version of this table is available in the online version of this article.}}
\label{tab:sv1quasar}
\end{deluxetable}

\section{Contaminants} \label{C}
\begin{figure}%
\centering 
\epsscale{1.2}
\plotone{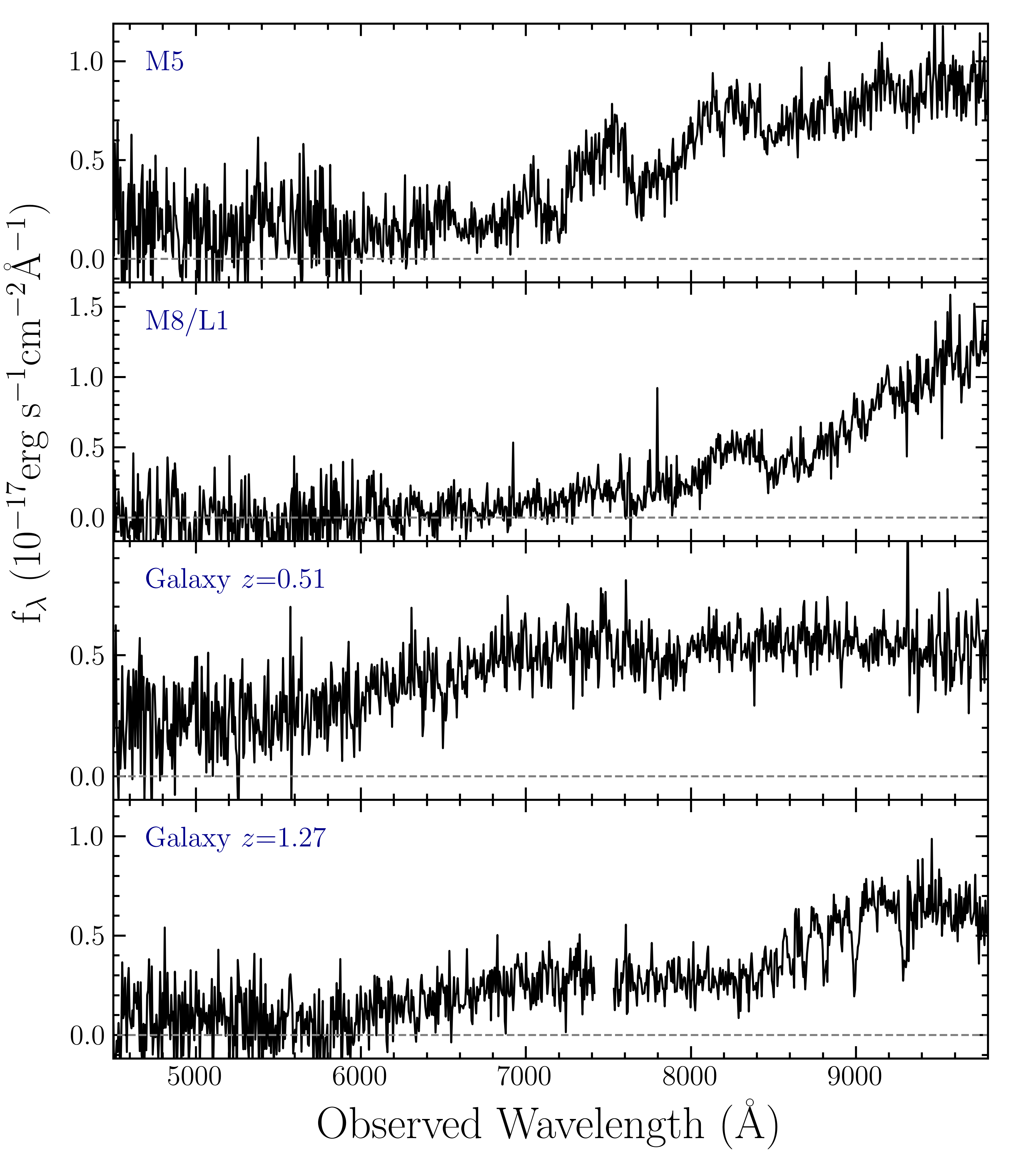} 
\caption{The example spectra of contaminants from our quasar survey, rebinned with seven pixels. The first two are M dwarfs and the other two are galaxies. They all have red colors and thus contaminate the quasar selection.}
\label{fig:contam}
\end{figure}

In Figure \ref{fig:contam} we show example spectra of the main contaminants, M/L/T dwarfs and red galaxies, of our high-redshift quasar survey. We identify them using SDSS templates. There is one object that can not be certainly identify as a late type M dwarf or an early type L dwarf based on the current optical spectrum. 

\end{document}